\documentclass[prb,twocolumn,longbibliography,superscriptaddress,nofootinbib,floatfix]{revtex4-2}
\usepackage{graphicx}
\usepackage{textgreek}

\usepackage{amsmath}
\usepackage{amssymb}
\usepackage{hyperref}
\usepackage[caption=false]{subfig}
\usepackage[capitalize]{cleveref}
\usepackage{braket}
\usepackage{mathtools}
\usepackage{interval}
\usepackage{bm}
\intervalconfig{soft open fences}

\AtBeginDocument{

}

\DeclareMathOperator{\tr}{tr}
\DeclareMathOperator{\Tr}{Tr}

\DeclarePairedDelimiter\abs{\lvert}{\rvert}
\newcommand{\lrvec}[1]{\overset{\text{\tiny$\bm\leftrightarrow$}}{#1}}

\crefname{section}{Sec.}{Sections}

\begin{document}
\title {Two-dimensional Stoner transitions beyond mean-field}
\author{Zachary M. Raines}
\affiliation{School of Physics and Astronomy and William I. Fine Theoretical Physics Institute, University of Minnesota, Minneapolis, MN 55455, USA}
\author{Andrey V. Chubukov}
\affiliation{School of Physics and Astronomy and William I.
    Fine Theoretical Physics Institute, University of Minnesota, Minneapolis, MN 55455, USA}
\begin{abstract}
We have previously shown that the Stoner instability 2D has unconventional behavior: it is strongly first order but features a susceptibility which diverges at the transition point.
Here, we analyze the Stoner transition for two-dimensional systems with spin and valley degrees of freedom, beyond mean field.
At low density, we show that in one-valley and isotropic two-valley systems the leading effect beyond mean-field theory is suppression of the Stoner instability.
In anisotropic two-valley systems, we show that, for larger anisotropy, the transition remains mean-field-like and retains its unconventional properties.
We discuss applications to AlAs quantum wells.
\end{abstract}

\maketitle

\section{Introduction}

For nearly the two decades since the experimental fabrication of graphene, there has been broad experimental interest in quasi- two-dimensional (2D) semiconductors whose low-energy theory hosts spin and valley degrees of freedom.
In this time there have been many proposals for how the valley iso-spin allows for new phenomena and applications, a field given the moniker ``valleytronics''~\cite{Bistritzer2010,Schaibley2016}  in analogy with the spintronic applications of older semiconductors~\cite{Zutic2004}.

Interest in systems with such valley and spin degrees of freedom only intensified when superconductivity was discovered in twisted bilayer graphene (TBG)~\cite{Cao2018,Cao2018a,Andrei2020} in close proximity to correlated phases, which may possess an isospin order~\cite{Tsvelik2023}.

Since then, several experiments have observed the formation of spontaneously iso-spin polarized states in other quasi-two-dimensional systems with spin and valley degrees of freedom, including non-twisted Bernal-stacked bilayer graphene (BBG), rhombohedral tri-layer graphene (RTG) in displacement field~\cite{Seiler2022,Zhou2022a,Zhou2021,DeLaBarrera2022,Seiler2023,Arp2023,Holleis2023,Zhang2023a,Blinov2023,*Blinov2023a} and AlAs quantum wells~\cite{Shayegan2006, Hossain2020,Hossain2021,Hossain2022}. These experiments triggered
an array
of theoretical studies of these systems~\cite{ghazaryan2021unconventional,chatterjee2022inter,Chichinadze2022,Chichinadze2022a,You2022,Szabo2022,Zang2022,Xie2023,Dong2023, Dong2023b,Szabo2022,Xie2023,Valenti2023,Calvera2024}.

Measurements near the onset of spin order in BBG, RTG, and AlAs hint at unconventional behavior:
approaching the transition from the unpolarized side, soft-bosonic excitations are observed e.g., a growing static iso-spin susceptibility in AlAs~\cite{Shayegan2006, Hossain2020};
however, the transitions themselves are strongly first order, with some number of iso-spin bands being completely depopulated
immediately below the transition~\cite{Zhou2021,Zhou2022a,DeLaBarrera2022,Arp2023,Holleis2023,Zhang2023a, Gunawan2006, Shayegan2006, Hossain2020,Hossain2021,Hossain2022}.

L. Glazman and the two of us (RGC) recently argued~\cite{Raines2024a,Raines2024b} that these unconventional features can be qualitatively reproduced by analyzing iso-spin Stoner transitions in 2D within
mean-field (Hartree-Fock) theory for a system of fermions with a repulsive interaction $U$ and isotropic dispersion.
We demonstrated that in a one-valley system with a parabolic dispersion, the Stoner transition at a critical $U=U_c$  is a first-order transition into a fully polarized ferromagnetic state, yet it is accompanied by the divergence of the magnetic susceptibility at $ U = U_c -0^{+}$ (see \cref{sec:mf} below).
For a two-valley system with SU(4) symmetry and parabolic dispersion, the Stoner instability is a first-order transition
into a fully spin and valley polarized quarter-metal state,
again accompanied by the divergence of an iso-spin susceptibility.
When SU(4) symmetry is broken, there is a cascade of two first-order transitions, the first one into a half-metal and the second one into a quarter-metal.
Like in other cases, each transition is accompanied by a divergence of the corresponding iso-spin susceptibility (\cref{sec:two-valley-mf}).

In this work, we extend the previous analysis by going beyond mean-field
(MF) and also including the anisotropy of the dispersion.
It has been argued earlier that at least in a single-valley system the existence of a
ferromagnetic
transition
beyond the MF level is not guaranteed
because particle-particle renormalizations suppress $U$ and may keep it below the level required for the Stoner instability~\cite{Kanamori1963,Irkhin2001,Ojajarvi2024}.
On the other hand, E.
Lieb has rigorously shown~\cite{Lieb1989} that a
ferromagnetic
transition does occur at a finite Hubbard $U$ for a single-valley fermionic system on the half-filled Lieb lattice\cite{Lieb1989}, and experiments on BBG, RTG, AlAs, and other materials did show a cascade of transitions, identified as
instabilities towards an iso-spin order.

From a general point of view, there are three possibilities beyond MF:
\begin{enumerate}
    \item[(I)]  A MF-like transition, perhaps at larger $U$,
    \item[(II)] A first-order transition to a polarized state, with no divergence of the susceptibility,
    \item[(III)] No transition.
\end{enumerate}
In this work, we obtain
the leading beyond-MF contributions to the 2D Stoner
susceptibility and the ground state energy at low electron density and attempt to distinguish between these three possibilities. We show that in both single-valley and
isotropic two-valley systems, the dominant effect beyond MF is a downturn renormalization of the interaction
that determines the Stoner instability, by logarithmically large contributions from the particle-particle channel.
Treating these logarithmic contributions within the ladder approximation, we find that they suppress the iso-spin susceptibilities \emph{and} make the iso-spin polarized state energetically unfavorable.
This clearly indicates that the MF approximation severely overestimates the system's ability to develop a Stoner instability.
Taken at a face value, these results are consistent with the case (III).
However, what actually happens at larger $U$ remains an open question because
the logarithmic approximation neglects subleading, non-logarithmic corrections in the particle-particle channel, which at large enough $U$ may invalidate the logarithmic approximation and allow an iso-spin susceptibility to diverge at some critical $U$ (this would be case I).
\footnote{An example of such behavior is pairing at a quantum-critical point in a metal and in the Yukawa-SYK model,  where ladder series of Cooper logarithms in the pairing vertex do not give rise to an instability, but once the coupling exceeds some critical value,
    subleading terms invalidate ladder summation, and the system becomes unstable towards pairing (see e.g.
    Refs.~\cite{Abanov2020,Hauck2020}).}
Also, RPA-type particle-hole corrections to the ground state energy may shift the energy balance in favor of
an iso-spin polarized state even if a corresponding iso-spin susceptibility remains finite (this would be case II).

For a two-valley system with an anisotropic quadratic dispersion, like the one in AlAs --- two bands centered at the $X$ and $Y$ points in the Brillouin zone (centers of the two valleys) with $\epsilon_{\mathbf{k},\tau} = (\eta^{\tau}k^{2}_{x}+\eta^{-\tau}k^{2}_{y})/2m$, where $\eta \neq 1$ and $\tau = \pm 1$ in the two valleys ---
SU(4) symmetry is broken already in the MF approximation.
Yet, within the MF, the degeneracy between valley polarization and intra-valley ferromagnetism remains, and the Stoner transition within MF is again a first-order transition into a fully polarized, spin/valley ordered quarter-metal state, accompanied by the divergence of an iso-spin susceptibility.
Beyond MF, we show in \cref{sec:two-valley-beyond} that the degeneracy between Stoner transitions into a ferromagnetic state and a valley polarized state is broken.
For the ferromagnetic interaction logarithmic corrections from the particle-particle channel are the same as in the isotropic case, and as long as logarithmic approximation is valid, there is no Stoner instability.
For the interaction that governs the Stoner transition into a valley polarized state, however, the strength of the logarithmic corrections depends on the anisotropy parameter $\eta$.
For large $\eta$, we find that the critical $U_c$, at which the valley polarization susceptibility diverges,
is comparable to the MF value.
In \cref{sec:gs},
we analyze the ground state energy beyond MF and argue that within the same logarithmic approximation, the system undergoes a first-order transition into fully valley-polarized state at exactly the same $U_c$.
This is scenario I: a MF-like Stoner transition.

We emphasize that our result is based on the summation of ladder series of logarithmical renormalizations from the particle-particle channel both for the susceptibility and for the ground state energy.
In a somewhat different approach, the authors of~[\onlinecite{Calvera2024}] summed up the series of RPA-type
ring
diagrams for the ground state energy.
They found first-order valley polarization and ferromagnetic transitions (in this order), not accompanied by the divergence of the corresponding susceptibilities (scenario II).
We compare the two calculations in \cref{sec:rpa}.

In what follows, we will reserve the designation \emph{Stoner} for instabilities where the iso-spin susceptibility diverges at the transition point.
In this language, MF transitions in a single-valley and a two-valley system are Stoner transitions, while beyond MF the three choices are a Stoner transition, a first-order transition not accompanied by the divergence of the corresponding susceptibility, or no transition at all.

In order to highlight the role played by valley physics, we first consider a single-valley system within and beyond MF, and then do the same analysis for a two-valley system.

\section{A single-valley system}
\label{sec:mf}

\begin{figure}
    \centering
    \includegraphics[width=0.5\linewidth]{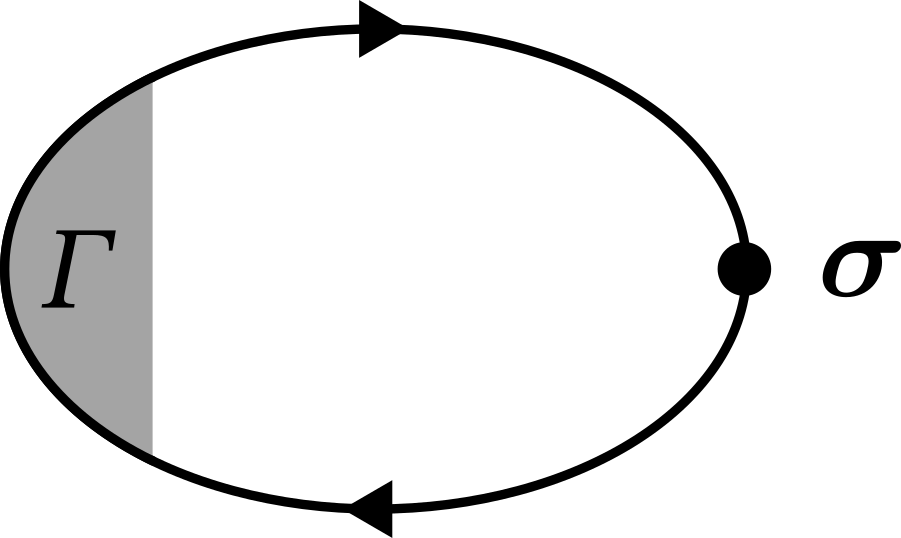}
    \caption{Full spin susceptibility.
        Solid lines are the full Green's function and the shaded vertex the fully renormalized spin vertex}
    \label{fig:spin-susc-full}
\end{figure}
\begin{figure}
    \centering
    \includegraphics[width=0.8\linewidth]{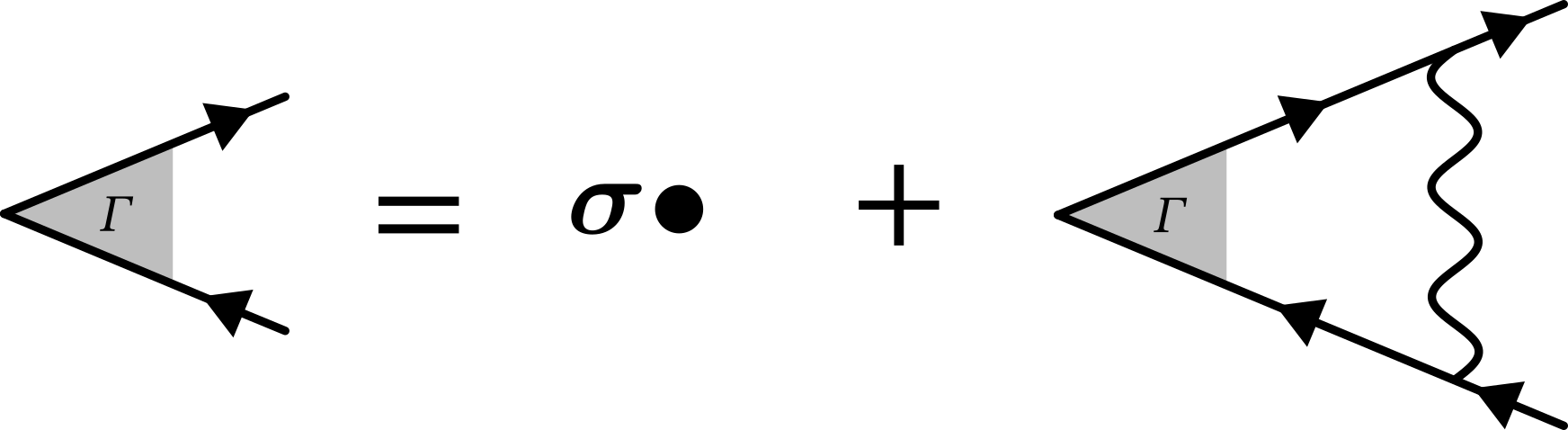}
    \caption{
        One-loop Bethe-Salpeter equation for the renormalized spin vertex.
    }
    \label{fig:hf-susc}
\end{figure}

We consider a model of two-dimensional spin-full fermions with repulsive contact interaction $U>0$
\begin{multline}
    \hat{H} = \sum_{\mathbf{k}\sigma}
    \left(\frac{k^{2}}{2m}-\mu_0\right)
    c^{\dagger}_{\mathbf{k}\sigma}c_{\mathbf{k}\sigma}\\
    + \frac{U}{2}\sum_{\mathbf{k}\mathbf{k}'\mathbf{q}\sigma\sigma'}
    c^{\dagger}_{\mathbf{k}+\mathbf{q}/2,\sigma}
    c^{\dagger}_{\mathbf{k}'-\mathbf{q}/2,\sigma'}
    c_{\mathbf{k}'+\mathbf{q}/2,\sigma'}
    c_{\mathbf{k}-\mathbf{q}/2,\sigma},
    \label{eq:stoner2d}
\end{multline}
where $\sigma$ indexes spin and $\mu_{0}$ is the chemical in absence of interactions.

Our goal is to obtain the spin susceptibility in the paramagnetic state and the ground state energy in the presence of a ferromagnetic order.
The paramagnetic spin susceptibility is the convolution of wo fermionic propagators and the fully renormalized spin vertex $\Gamma_k$ (see~\cref{fig:spin-susc-full})
\begin{equation}
    \chi =-2T\sum_{k} G^{2}_{k} \Gamma_k
    \label{eq:chi-1loop_1}
\end{equation}
where $k=(\mathbf{k}, \epsilon_{n})$ is the internal momentum and frequency,
and $\Gamma_k$ is the solution of the Bethe-Salpeter (BS) equation
\begin{equation}
    \Gamma_k\hat{\bm{\sigma}} = \hat{\bm{\sigma}}- T \sum_{p}  G^{2}_{p} \Gamma_p
    \lrvec{S}_{k,p} \hat{\bm{\sigma}},
    \label{eq:bs-1loop_1} \end{equation} and
where $\lrvec{S} (k,p)$ is the fully renormalized irreducible four-point function, which generally has some spin structure.

To calculate the ground state energy one can follow the variational Luttinger-Ward approach~\cite{Luttinger1960}.
The Luttinger-Ward free energy (ground state energy at $T=0$)
\begin{equation}
    E_{LW}[\hat{G}] =
    -\ln
    \det[-\hat{G}^{-1}]
    - \tr[\hat{\Sigma}\hat{G}] + \Phi[\hat{G}].
    \label{eq:E-LW-var}
\end{equation}
is a functional of the full Green's function
\begin{equation*}
    \hat{G} =
    \begin{pmatrix}
        \hat{G}_{\uparrow} & 0                    \\
        0                  & \hat{G}_{\downarrow}
    \end{pmatrix}
\end{equation*}
and
the self-energy $\hat{\Sigma} = {\hat G}_0^{-1} -{\hat G}^{-1}$ is defined by $\hat{\Sigma}\equiv\delta\Phi/\delta \hat{G}$.
The Luttinger-Ward functional $\Phi[\hat{G}]$ is obtained by collecting closed-loop diagrams in the loop expansion with the full Green's functions.
Working in terms of the Luttinger-Ward functional has the benefit of automatically producing a self-consistent and conserving approximation~\cite{Luttinger1960,Baym1961a,Baym1962}.

We first review this procedure in the MF approximation and then move beyond MF.\@

\subsection{Spin susceptibility in MF}

We start with the
normal state
susceptibility.
The MF approximation uses the bare $G_{0;k}$ instead of $G_k$ in
(\ref{eq:bs-1loop_1})
and approximates $S_{k,p}$ by the bare interaction $U$.
This makes $\Gamma$ independent of $k$.
We then have
\begin{equation}
    \chi_{MF} =-2T\sum_{k} G^{2}_{0;k} \Gamma
    \label{eq:chi-1loop}
\end{equation}
and
\begin{equation}
    \Gamma\hat{\bm{\sigma}} = \hat{\bm{\sigma}}- T U \sum_{p}  G^{2}_{0;p} \Gamma \hat{\bm{\sigma}},
    \label{eq:bs-1loop}
\end{equation}
At low temperatures \begin{equation}
    \lim_{T\to0} T \sum_{k} G^{2}_{0;k}
    = -\nu,
\end{equation}
and the solutions of \cref{eq:chi-1loop,eq:bs-1loop} can then be
explicitly written as
\begin{equation}
    \Gamma = \frac{1}{1 - \nu U},\quad \chi_{MF} = \frac{2\nu}{1 - \nu U}.
    \label{eq:chi-1loop-explicit}
\end{equation}
The renormalized spin vertex and
the spin susceptibility diverge at $\nu U = 1$,
signaling the ferromagnetic instability of the normal state.

\subsection{Ground state energy in MF}
\label{sec:gs-mf-spin}

\begin{figure}
    \centering
    \includegraphics[width=0.6\linewidth]{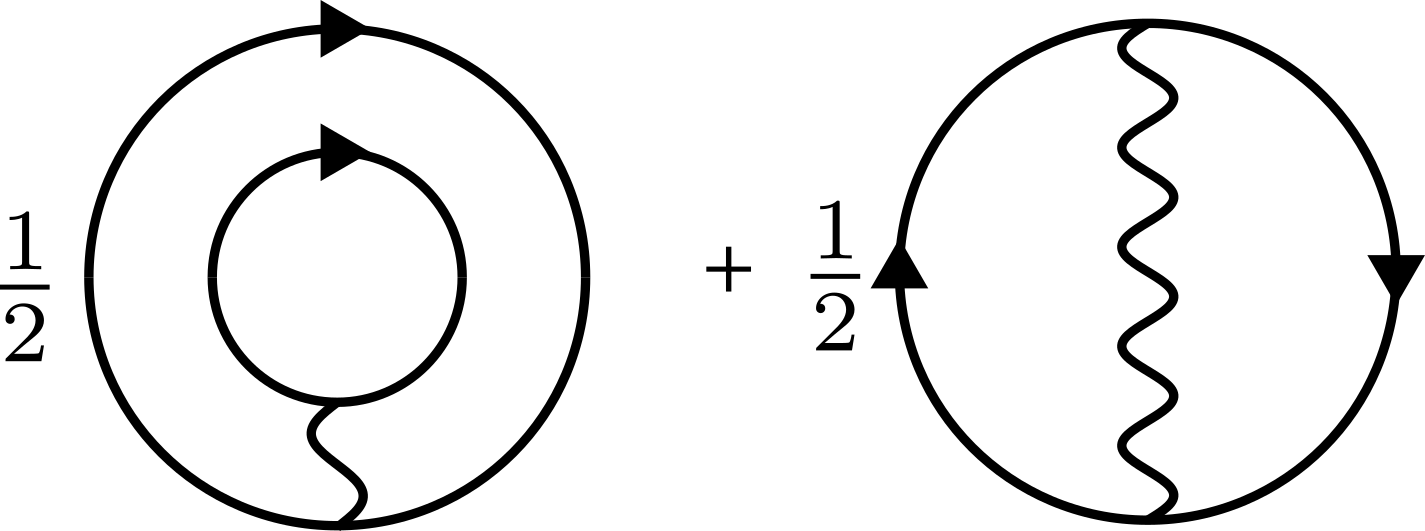}
    \caption{MF approximation to the Luttinger-Ward functional $\Phi$.
        Solid fermionic lines represent the matrix Green's functions with spin index $\sigma$.
        The wavy line represents the bare Hubbard interaction $U$.         Note that the diagrammtic rules for the Luttinger-Ward functional include an extract minus sign compared to the usual.
        \label{fig:lwMF}}
\end{figure}

Within the same approximation, one should use only Hartree and Fock diagrams for the Luttinger-Ward functional $\Phi [\hat{G}]$, see~\cref{fig:lwMF},
\begin{equation}
    \Phi_{MF}[\hat{G}]\approx
    \frac{1}{4}UT^{2}\sum_{kk'}
    \left(
    \Tr[\hat{G}_{k}]\Tr[\hat{G}_{k'}]- \Tr[\hat{G}_{k}\hat{G}_{k'}].
    \right).
    \label{eq:lwMF-spin}
\end{equation} Because the corresponding fermionic $\hat \Sigma$ reduces to a constant, it can be absorbed into the renormalization of the chemical potential from $\mu_0$ to the actual $\mu$, related to fermionic density, and a constant shift of the spin-bands relative to each other, related to the spin polarization.
One can then easily verify that this approximation is equivalent to evaluating kinetic and potential energies with free-fermion
propagators in the presence of spin polarization
and keeping only Hartree and Fock diagrams for the potential energy.
One can also verify that the MF expression for $\chi_{MF}$ is obtained by applying a magnetic field and differentiating the ground state energy, obtained this way, twice over the field.

\begin{figure}
    \centering
    \includegraphics[width=0.9\linewidth]{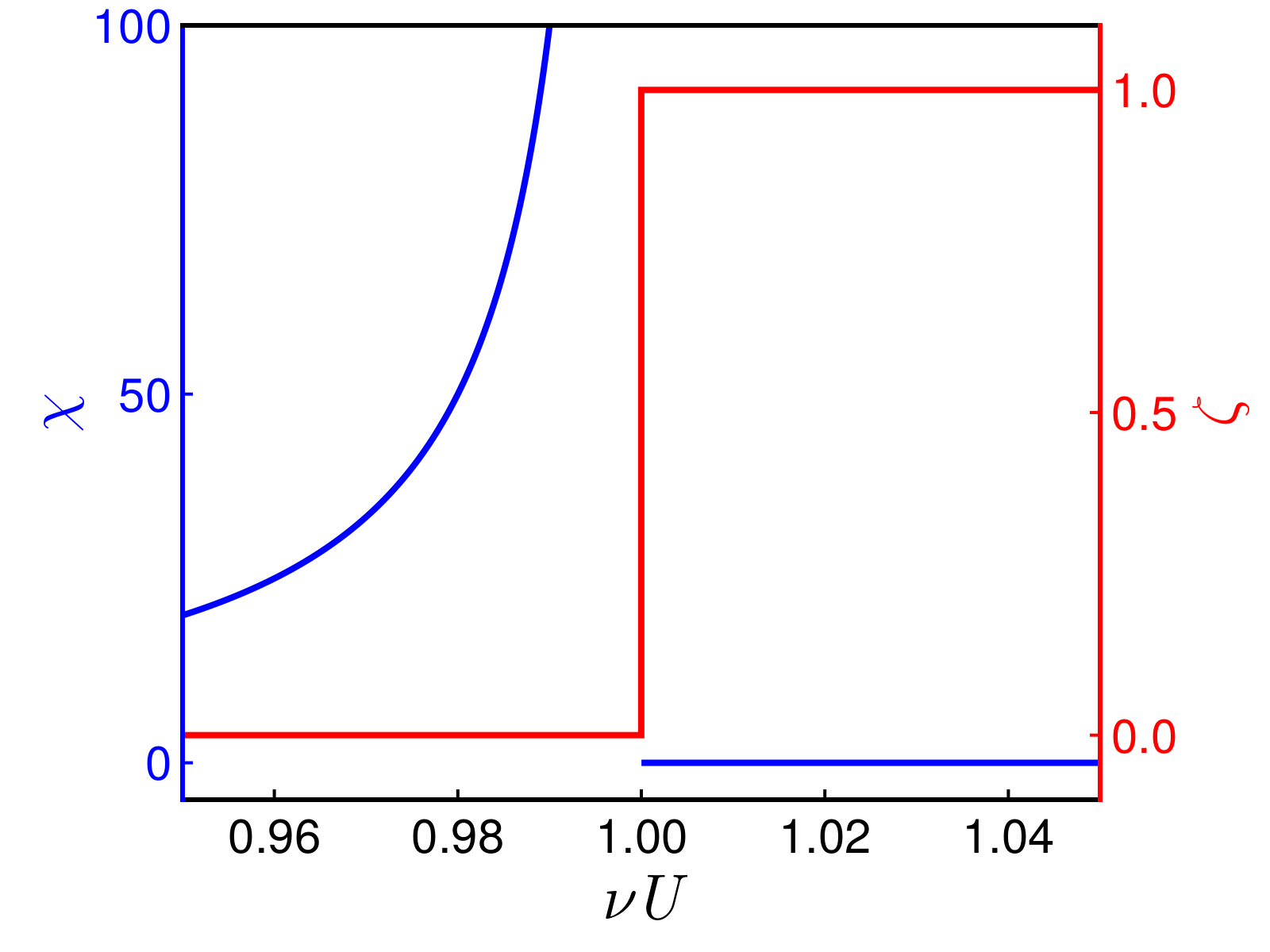}
    \caption{(Color online)
        Spin susceptibility $\chi$ (blue) and polarizaton $\zeta$ (red) as function of the dimensionless interaction strength $\nu U$.
        At $\nu U = 1$ there is a first order transition into a fully spin-polarized state, and the spin-susceptibility diverges as this point is approached from the normal state.
        \label{fig:chi-and-zeta}}
\end{figure}

In 2D, the MF ground state energy turns out to be
a quadratic function of the dimensionless spin polarization $\zeta \equiv (n_{\uparrow}-n_{\downarrow})/n$~\cite{Raines2024a,*Raines2024b}
\begin{equation}
    E_{MF}  =
    E_{N}  + \frac{n^{2}}{2\nu}
    \left(1 - \nu U\right)\zeta^{2},
    \label{eq:Emf}
\end{equation}
where $E_{N}$ is the energy of the normal state,
$n = n_{\uparrow}+n_{\downarrow}$ is the fixed total density, and $\abs{\zeta}\leq1$.
There are no terms of higher order in $\zeta$.
This form of $E_{MF}$ indicates that
the system undergoes a
first order transition to a fully spin-polarized state $\abs{\zeta}=1$
at $\nu U = 1$, i.e., at exactly the same point where the static spin-susceptibility \cref{eq:chi-1loop-explicit} diverges.
We illustrate this in \cref{fig:chi-and-zeta}.
We emphasize that this behavior is a consequence of the fact that the energy is exactly quadratic in the polarization, $\zeta$.
~\footnote{
    For a generic isotropic power-law dispersion $\epsilon_k \propto k^{2\alpha}$,
    a first order transition of Stoner type
    develops for both $k^2$ and $k^4$ dispersions ($\alpha =1$ and $\alpha =2$).
    For intermediate $1 < \alpha <2$, a first-order transition occurs before the susceptibility diverges, but the difference between the critical $U$ where this happens and where the susceptibility would diverge is
    less than 2\%~\cite{Raines2024a,*Raines2024b}, i.e., to a truly good approximation, a first-order Stoner transition holds for all power-law dispersions with $1\leq \alpha \leq 2$.}.

\subsection{Spin susceptibility beyond MF}
\label{sec:spin}

\begin{figure}
    \centering
    \includegraphics[width=\linewidth]{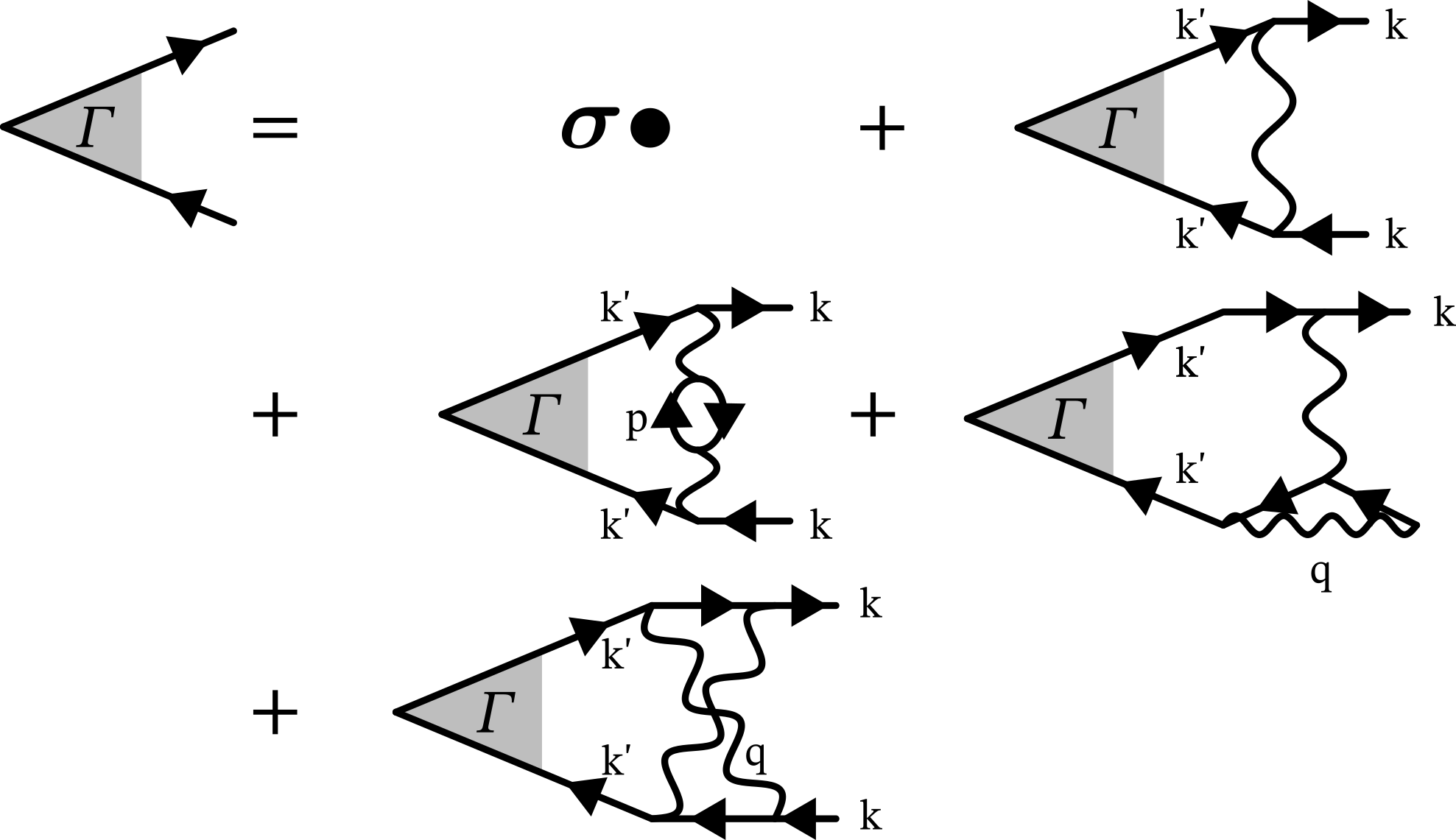}
    \caption{Bethe-Salpeter equation for the renormalized spin vertex at two-loop order.
        The diagrams in the second and third row represent corrections to the self-consistent
        MF
        result.}
    \label{fig:lowest-correction}
\end{figure}

To go beyond MF, we first ask what the most important contribution is at the next order, i.e., two-loop.
In doing so, we will see that for the low-density limit, one of the contributions is universally larger than the other.
Concretely,
let us look at contributions obtained by keeping irreducible four-point function $\lrvec{S}_{k,p}$ to two loops.
At two-loop order there are three contributions, shown in \cref{fig:lowest-correction}.

For the Hubbard model, the polarization insertion and vertex correction diagrams (the ones in the second line in \cref{fig:lowest-correction}) cancel, leaving just the crossed diagram.
Even so, let us look at the behavior of each of two-loop diagram.
For all diagrams the external fermionic momentum is near the Fermi surface $\mathbf{k}\sim\mathbf{k}_{F}$.
Additionally, the factor of $G^{2}_{0;k'}$ from the lines adjacent to the vertex $\Gamma$ means that the dominant contribution comes from
$\mathbf{k}' \sim \mathbf{k}_{F}$.
For the polarization insertion and vertex correction diagrams, the restrictions on $k$ and $k'$ are
sufficient to restrict other internal fermionic momenta to the vicinity of the Fermi surface.
This leads to regular corrections to the vertex, of order $\nu U$, as one can easily verify.
The story is qualitatively different for the crossed diagram.
Here we have
\begin{equation}
    \Gamma_\text{crossed} = U^{2} T^{2} \smashoperator{\sum_{k',q}} G^{2}_{0;k'}
    G_{0;k-q}
    G_{0;k'+q}\Gamma.
\end{equation}
While the Matsubara sums restrict $\mathbf{k},\mathbf{k'}$ to be near the Fermi surface, the integration over $\mathbf{q}$ is
unconstrained:
because this diagram involves a particle-particle process, all we require is that the intermediate particles both be above or below the Fermi surface.
This has strong implications for low-density systems, i.e., those in which $k_{F}$ can be considered small compared to the upper cutoff of the model, $\Lambda$.
There is a contribution from $q \lesssim 2k_{F}$ leading to a regular $O(\nu U)$ term, but there is also a contribution coming from $k_F \ll q \ll \Lambda$.
To evaluate this piece we can, at leading order, neglect the fermionic momenta $\mathbf{k},\mathbf{k}'$ and
the corresponding frequencies $\epsilon_{n},\epsilon_{n'}$ compared to $\omega_{m},\mathbf{q}$ and approximate
\begin{equation}
    \Gamma_\text{crossed} \approx U^{2} T^{2} \smashoperator{\sum_{k'q}} G^{2}_{0;k'}
    G_{0;-q}
    G_{0;q}\Gamma
    = -U^{2}\Pi^{ph}\Pi^{pp}\Gamma,
    \label{eq:gamma-crossed-spin-only}
\end{equation}
where we have defined the particle-particle bubble
\begin{equation}
    \Pi^{pp} \equiv T \sum_{\omega_{m}}\sum_{|\mathbf{q}| \gg k_{F}}
    G_{0;-q}
    G_{0;q}.
\end{equation} In the low temperature limit, $\Pi^{ph}=\nu$ as noted above, while
\begin{equation}
    \Pi^{pp}  = -\int \frac{d\omega_{m}}{2\pi}\int_{q_{\text{min}}}^{\Lambda}\frac{q dq}{(i\omega_{m})^{2}-\epsilon^{2}_{\mathbf{q}}}.
    \label{eq:pi_pp}
\end{equation}
This has a familiar form, expected for the particle-particle polarization,
but there is a lower cutoff at $q_{min} \sim k_F$
$\mathbf{k},\mathbf{k}'$ can no longer be neglected at smaller $q$.
The upper cutoff $\Lambda$ is determined by the UV properties of the model and may take different forms depending on the nature of the underlying lattice model.
Since we are considering a Hubbard-like model here, $\Lambda$ is of order the Brillouin zone size.
We now notice that the integral in \cref{eq:pi_pp} is logarithmically singular.
This is not a Cooper logarithm as \cref{eq:pi_pp} is only valid at $q > k_F$, but rather a 2D-specific logarithm associated with the logarithmic behavior of the scattering amplitude.
Evaluating the integral in~\cref{eq:pi_pp} to logarithmic accuracy, we find,
\begin{equation}
    \Pi^{pp}  \approx -\nu\int \frac{d\omega_{m}}{2\pi}\int_{k^{2}_{F}/(2m)}^{\Lambda^{2}/(2m)}
    \frac{d\xi}{(i\omega_{m})^{2}-\xi^{2}}=
    \frac{\nu}{2} \ln \frac{\Lambda}{k_{F}}.
\end{equation}
For systems with
$k_{F}\ll \Lambda$ the logarithm is parametrically large.
\begin{figure}
    \centering
    \includegraphics[width=\linewidth]{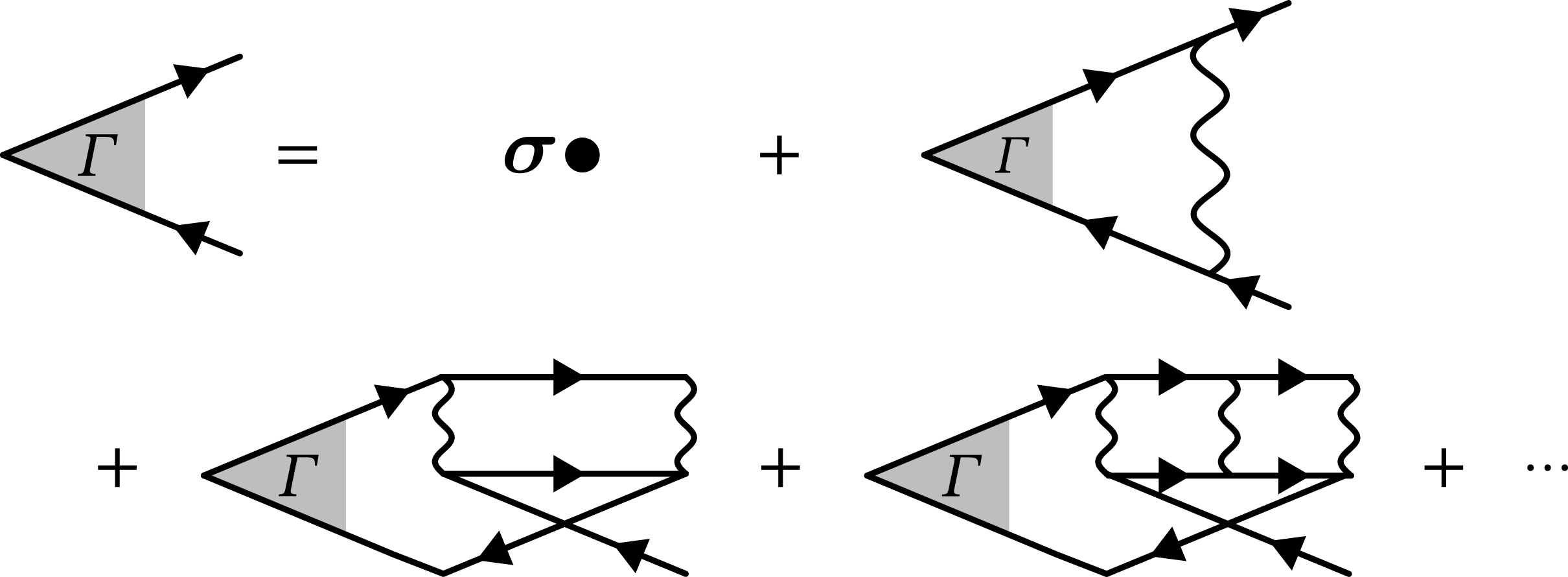}
    \caption{Leading contribution to the renormalized spin vertex in the low-density limit.
        The dominant contribution comes from particle-particle ladders which can, to logarithmic accuracy, be resummed into an effective contact interaction $\tilde{U}$, \cref{eq:tilde-U}.
        \label{fig:spin-bs-ladder}
    }
\end{figure}
We can thus approximate the \emph{entire} two-loop Bethe-Salpeter equation
as
\begin{equation}
    \Gamma_{2\text{loop}}  = 1 + \nu U \Gamma_{2\text{loop}} - \frac{1}{2}(\nu U)^{2} \ln \frac{\Lambda}{k_{F}} \Gamma_{2\text{loop}},
\end{equation}
where the terms on the right-hand side come from the bare, MF, and crossed contributions respectively.
The associated two-loop susceptibility is
\begin{equation}
    \chi_{2\text{loop}} = \frac{2\nu}{1 - \nu U + \frac{1}{2}(\nu U)^{2} \ln \frac{\Lambda}{k_{F}}}.
    \label{eq:2loop}
\end{equation}
We see from \cref{eq:2loop} that the leading effect at two-loop order is suppression of the Stoner instability due to particle-particle contributions away from the Fermi surface.
Furthermore,  $\chi_{2\text{loop}}$  from \cref{eq:2loop} does not diverge for any value of $U$.

With the insights from two-loop order we can go further.
Note that at $n$-loop order, there will be a contribution to the Bethe-Salpeter equation from the maximimally crossed diagram,
involving $n-1$ independent particle-particle terms that goes as $\ln(\Lambda/k_{F})^{n-1}$
Any other diagram of the same order in $U$ contains at most $\ln(\Lambda/k_{F})^{n-2}$.
Summing only these leading contributions to all orders (see~\cref{fig:spin-bs-ladder}), we find, to logarithmic accuracy,
\begin{equation}
    \Gamma = 1 + \nu U\Gamma
    - \frac{\frac{1}{2}(\nu U)^{2} \ln \frac{\Lambda}{k_{F}}}{
        1 + \frac{1}{2}\nu U \ln \frac{\Lambda}{k_{F}}
    }\Gamma.
\end{equation}
This is readily solved for
\begin{equation}
    \Gamma = \frac{1}{1-\tilde{U}{\nu}},\quad
    \tilde{U}  = \frac{U}{1+\frac{1}{2} \nu U \ln \frac{\Lambda}{k_{F}}}.
    \label{eq:tilde-U}
\end{equation}
Within the regime $\nu U <1$,
where the logarithmic expansion is controlled we have that $\tilde{U}\nu < 1$ and thus the susceptibility does not
diverge.

The result may intuitively be understood by noting that in the low density limit each
two-particle
collision can be considered independent from the other fermions.
Thus, the leading effect may be captured by replacing the bare interaction with the two-particle scattering amplitude,
which is logarithmically singular in 2D~\cite{Lifsic2006,galitskii1958energy,fetter2012quantum}.

\subsection{Ground state energy beyond MF}
\label{sec:gs-spin}
\begin{figure}
    \centering
    \includegraphics[width=\linewidth]{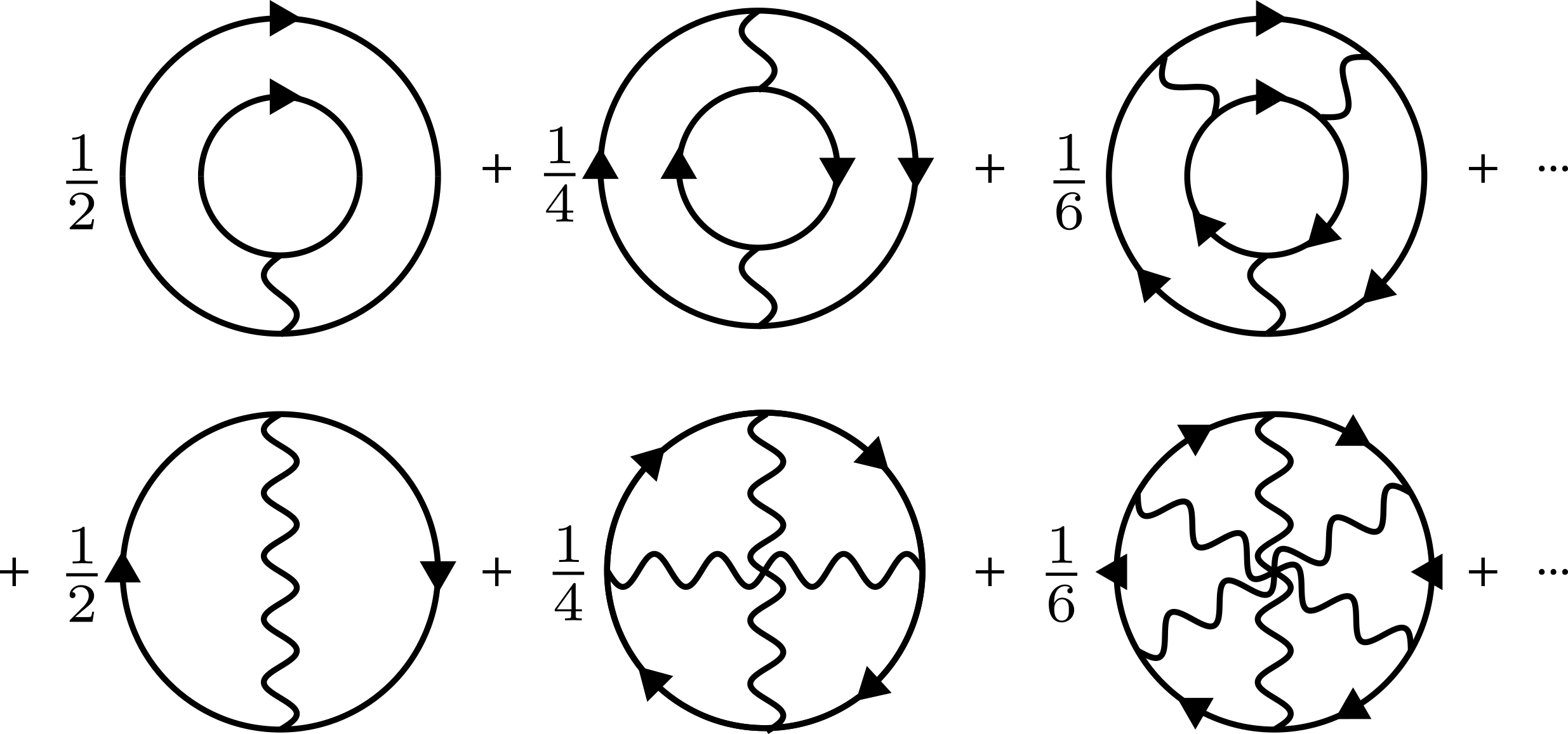}
    \caption{Leading contributions to the Luttinger Ward functional in the low-density limit $k_{F} \ll \Lambda$.,
        Solid fermionic lines represent matrix Green's functions with spin index $\sigma$.
        The MF approximation corresponds to keeping the first diagram of each line.
        \label{fig:lw-ladder}}
\end{figure}

\begin{figure}
    \centering
    \vspace{1em}
    \includegraphics[width=0.9\linewidth]{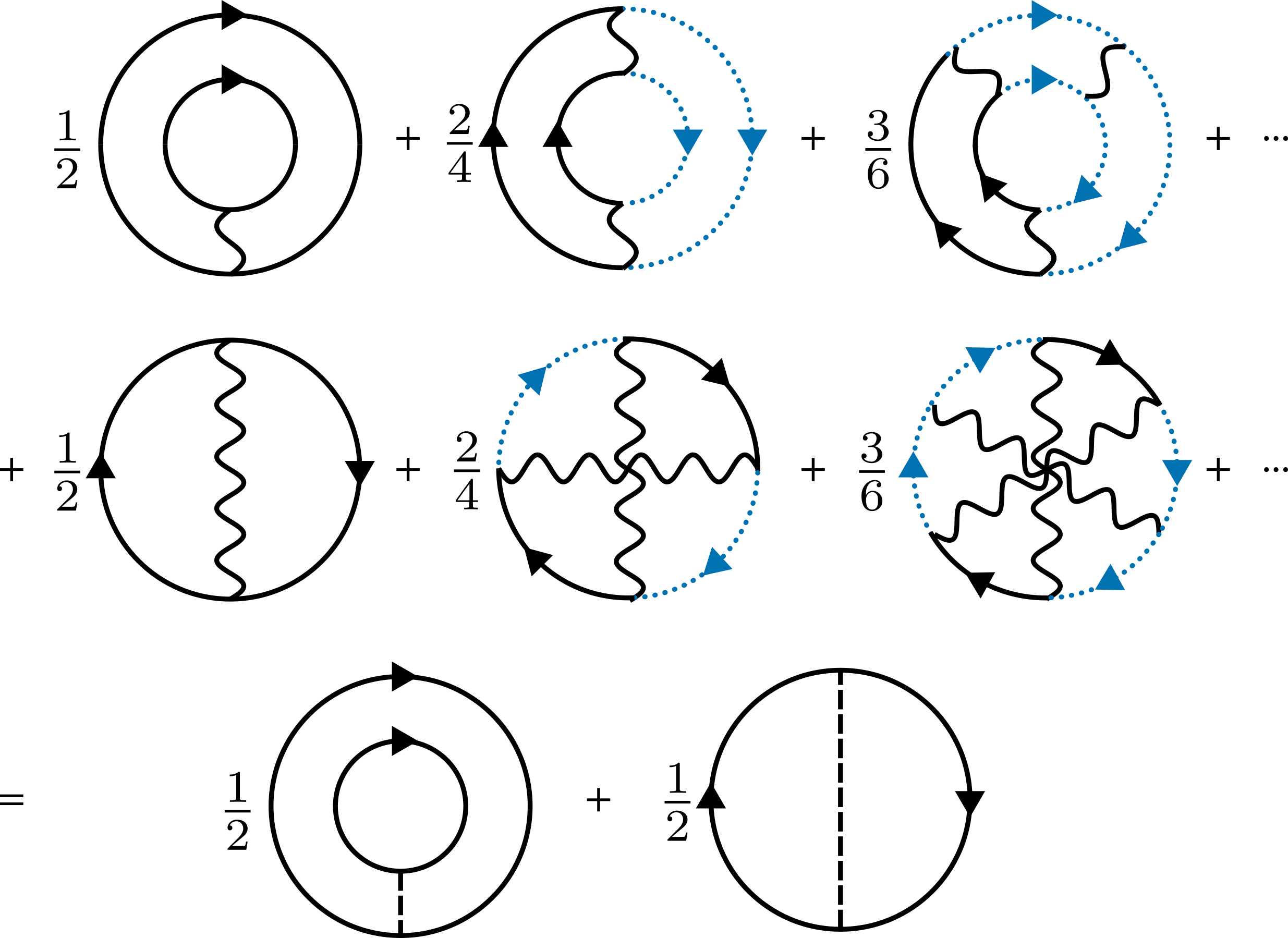}
    \caption{Reduction of the Luttinger Ward functional to an effective Hartree-Fock theory with logarithmic accuracy.
        The blue dotted lines represent fermions away from the Fermi surface.
        The combinatoric factors associated with ways of choosing two out of $2l$ lines to be near the Fermi surface cancel the factor of $1/l$ for each diagram.
        The effective interaction, represented by the dashed interaction line, is then obtained by summing particle-particle ladders in the usual way.
        \label{fig:lw-effective-MF}}
\end{figure}

To establish the ground state, we need to obtain the energy as a function of the spin polarization.
We will do so by computing the Luttinger-Ward functional within the same logarithmic approximation that we used for the susceptibility.
The
Luttinger-Ward functional containing the leading logarithmic terms at each order
is shown in \cref{fig:lw-ladder}.
Note that as is usual with LW approach, the Green's function lines represent the full Green's function $\hat{G}^{-1} = \hat{G}^{-1}_{0} - \hat{\Sigma}$, which must be solved for self-consistently~\footnote{The form of the variational Luttinger-Ward energy \cref{eq:E-LW-var} guarantees that minimizing with respect to $\hat{G}$ produces a self-consistent solution.}, and not the normal state Green's function $\hat{G}_{0}$.
The MF approximation~\cref{fig:lwMF} corresponds to keeping the first diagram in each of the two lines in \cref{fig:lw-ladder}.

Given the preceding discussion, the leading contribution to each diagram comes from when all but two pairs of particles are away from the Fermi surface, and the remaining particles are near the Fermi surface
(see \cref{fig:lw-effective-MF}), as these terms come with the maximum logarithmic enhancement at each loop order.
At this level, the combinatoric factors associated with ways of choosing $2$ out of $2l$ lines to be near the Fermi surface in an $l-$loop diagram cancel the overall factor of $1/l$, and
the Luttinger-Ward functional
\emph{retains the same form as in MF}, \cref{eq:lwMF-spin},
but with the
effective interaction $\tilde{U}$.
In explicit form,
\begin{equation}
    \Phi_{\log}[\hat{G}]\approx
    \frac{1}{4}\tilde{U}T^{2}\sum_{kk'}
    \left(
    \Tr[\hat{G}_{k}]\Tr[\hat{G}_{k'}]- \Tr[\hat{G}_{k}\hat{G}_{k'}]
    \right).
    \label{eq:lw-effective-MF-spin}
\end{equation}
where $\Tr$ is with respect to spin.
However, as noted above, as long as the logarithmic approximation holds $\tilde{U}\nu<1$ so there is no Stoner susceptibility within this regime.

\subsection{Discussion}
The general suppression of the Stoner
susceptibility
due to particle-particle ladders was discussed in the '60s by~\textcite{Kanamori1963}. At a face value, \cref{eq:tilde-U} implies that there is no Stoner transition (using our designation of \emph{Stoner transition}) in 2D.
We caution, however, that it is entirely possible that a Stoner transition actually
occurs but is pushed to larger values of $U$ where the logarithmic approximation breaks down.
Mathematically, this corresponds to the fact that while at each order in $\nu U$, the ladder diagrams are largest, the series as a whole may not be absolutely convergent.
Furthermore, there are systems for which it can be rigorously shown that a Stoner transition does occur, e.g., the
half-filled Hubbard model on the Lieb lattice\cite{Lieb1989}.

Given the uncertainty, we are left with the following three possibilities, which we already listed in the Introduction:
\begin{enumerate}
    \item[(I)] There is a transition of Stoner type, where the susceptibility diverges, outside the applicability of our resummation scheme,
    \item[(II)] There is a first-order transition to a polarized state
          with no divergence of the susceptibility,
    \item[(III)] There is no transition to a polarized state for any $U$.
\end{enumerate}
While that is all that can be theoretically said about the spin Stoner instability
in a single-valley system
within our treatment, we will see in the following sections that it is possible to make sharper predictions
for a
two-valley system.

\section{
  Two-valley system}
\label{sec:two-valley-mf}

We now consider how the story is modified in the presence of both spin and valley degrees of freedom.
This is of particular interest for iso-spin instabilities in quasi-two-dimensional multi-valley systems such as
BBG, RTG, and AlAs quantum wells.

\begin{figure}
    \centering
    \includegraphics[width=0.6\linewidth]{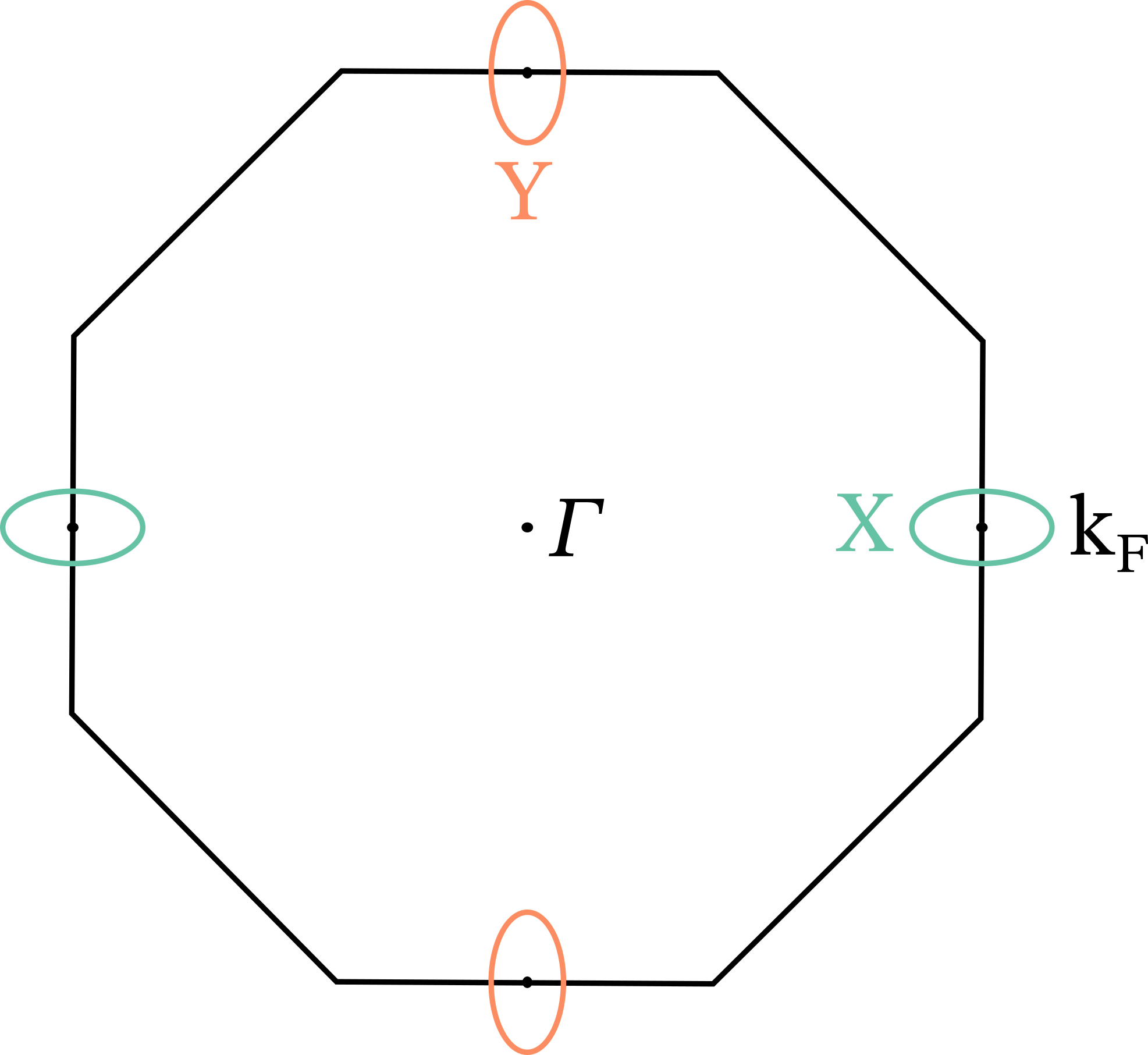}
    \caption{(Color online) Schematic depiction of the low energy bands
        of AlAs in the Brillouin zone.
        The two valleys correspond to regions near the $X$ (green) and $Y$ (orange) points in the Brillouin zone.
        There are spin degenerate ellipsoidal Fermi surfaces centered at
        $X$ and $Y$, related by a $C_{4}$ rotation.
        \label{fig:brillouin}
    }
\end{figure}

For definiteness, we consider a two-valley model appropriate for AlAs.
The low-energy description of this system has all of the necessary components while remaining relatively simple.
Bulk AlAs has a low-energy manifold consisting of ellipsoidal pockets near three
inequivalent points in the Brillouin zone (three valleys).
When confined along one of the periodic axes, e.g., $[001]$, and placed atop a GaAs substrate, the energy spectrum splits between lower-energy excitations in the two in-plane $X$ and $Y$ valleys (two pockets centered at $X =[100]$, and $Y =[010]$), and
higher-energy excitations in the out-of-plane $Z$ valley~\cite{Shayegan2006}.
The 2D dispersion around $X$ and $Y$ can be well approximated as quadratic with anisotropic band mass, leading to elliptical Fermi surfaces, see \cref{fig:brillouin}.

We follow~[\onlinecite{Shayegan2006,Hossain2020,Valenti2023,Calvera2024}] and model the low-energy physics of
AlAs by
\begin{equation}
    \hat{H}
    = \sum_{\mathbf{k},\sigma,\tau} \epsilon_{\mathbf{k},\tau}c^{\dagger}_{\mathbf{k}\tau\sigma}c_{\mathbf{k}\tau\sigma}
    + \hat{H}_{\text{int}}.
\end{equation}
where $\sigma$ and $\tau$ are spin and valley degrees of freedom, respectively, and
the valley dependent dispersion is
\begin{equation}
    \epsilon_{\mathbf{k},\tau} = \frac{\eta^{\tau}k^{2}_{x}+\eta^{-\tau}k^{2}_{y}}{2m}
    -\mu_{0}
\end{equation}
where the $\tau=+$ and $\tau=-$ valleys are centered, respectively, at the $X$ and $Y$ points.
The degree of anisotropy is set by the parameter $\eta$.
We will assume, without loss of generality, that $\eta\geq1$, but all results are manifestly invariant with regard to $\eta \to \eta^{-1}$.

\begin{figure}
    \centering
    \includegraphics[width=0.9\linewidth]{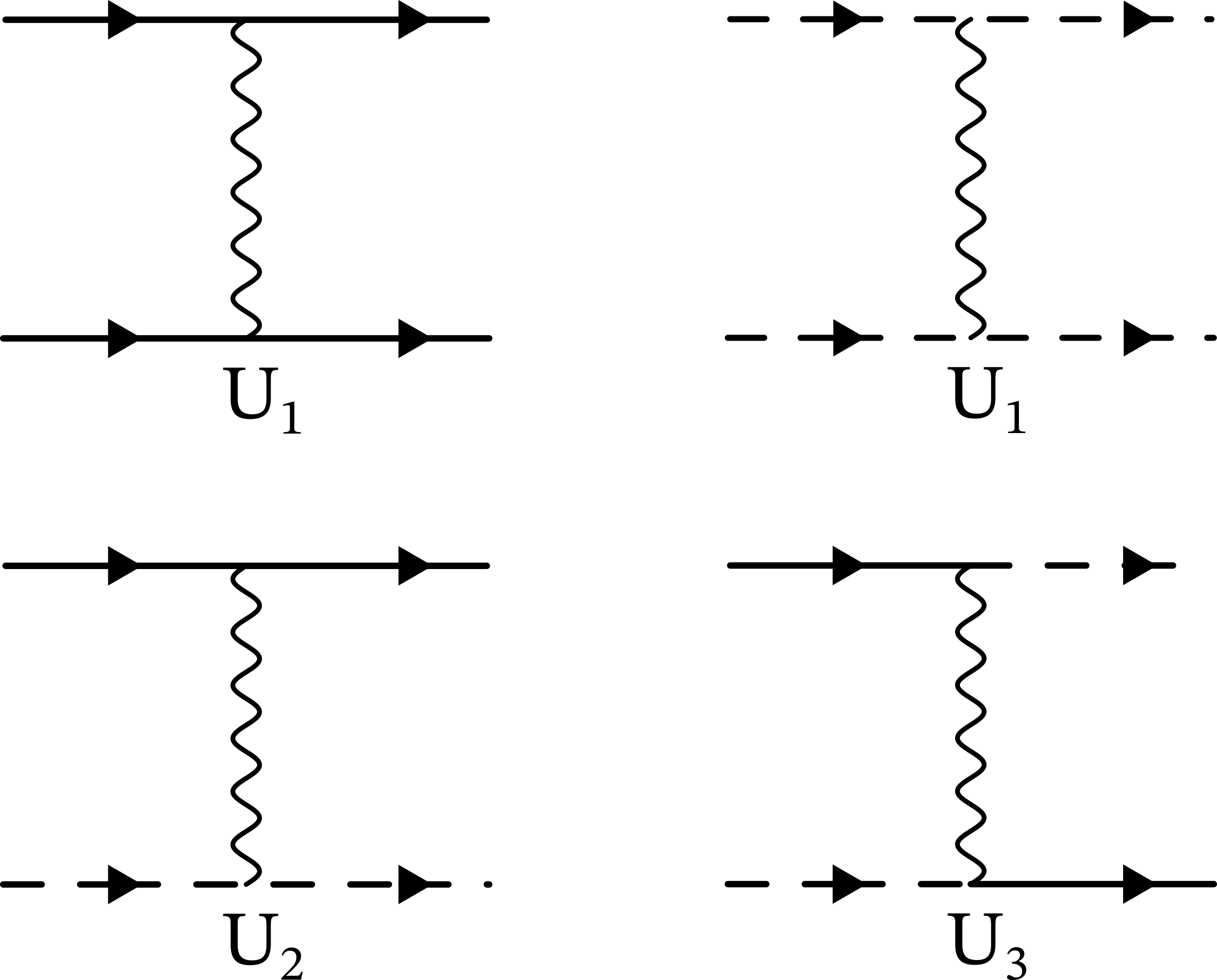}
    \caption{Density-density interactions
        in a multi-valley system.
        Solid lines represent fermions in the $\tau=+$ valley while dashed lines represent fermions in the $\tau=-$ valley.
        Independent frequency/momenta are labeled.
        \label{fig:interactions}}
\end{figure}

The interaction Hamiltonian $\hat{H}_{\text{int}}$
consists of three short ranged interactions, depicted in \cref{fig:interactions}:
interaction between densities in the same valley $U_{1}$, interaction between densities in different valleys $U_{2}$, and scattering from one valley to another $U_{3}$.
All three interactions arise from the screened Coulomb interaction at the appropriate momentum, $\mathbf{q}\sim 0$ for $U_{1},U_{2}$ and $\mathbf{q} \sim |\mathbf{X}-\mathbf{Y}|$ for $U_{3}$.
We consider the case where the Fermi momenta are much smaller than the separation between valleys $k_{F\tau}\ll|\mathbf{X}-\mathbf{Y}|$,  so that $U_{1}\approx U_{2}\gg U_{3}\approx 0$,
and neglect $U_3$~\footnote{As we will see later, it is important whether the Coulomb interaction is screened by the system itself our some other component of the setup, e.g., a nearby gate.}.
The interaction Hamiltonian then takes the form
\begin{multline}
    \hat{H}_{\text{int}}  = \frac{1}{2}   \sum_{\mathbf{k},\mathbf{k}',\mathbf{q},\tau\tau'\sigma\sigma'}
    (U_{1}\delta_{\tau\tau'} + U_{2}\tau^{1}_{\tau\tau'})\\
    \times
    c^{\dagger}_{\mathbf{k}+\mathbf{q}/2,\tau\sigma}
    c^{\dagger}_{\mathbf{k}'-\mathbf{q}/2,\tau'\sigma'}
    c_{\mathbf{k}'+\mathbf{q}/2,\tau'\sigma'}
    c_{\mathbf{k}-\mathbf{q}/2,\tau\sigma}
    \label{eq:Hint}.
\end{multline}

\subsection{Within MF}
\label{sec:2valley-mf}

Five types of spin/valley order can develop in this model within the MF approximation~\cite{Chichinadze2022a,Dong2023b}.
They are one-component valley-polarization (VP) ($\tau_z \sigma_0$),
three-component ferromagnetism (FM) ($\sigma_{x,y,z}\tau_0$),
three-component staggered ferromagnetism (SFM) ($\sigma_{x,y,z}\tau_{z}$),
two-component inter-valley coherence (IVC)
($\tau_{x,y} \sigma_0$),
and six-component spin inter-valley coherence (sIVC) ($\sigma_{x,y,z}\tau_{x,y}$).
For $\eta = 1$ and $U_{1}=U_{2}$, the model is SU(4) symmetric and the MF susceptibilities for all orders are
identical.
Analogous to the calculation of \cref{sec:mf} we find
\begin{equation}
    \chi^{SU(4)}_{MF}  = \frac{4\nu}{1-\nu U}.
\end{equation}
The
MF
ground state energy of SU(4)-symmetric model can be compactly written in a form similar to \cref{eq:Emf}:
\begin{equation}
    E_{MF}  = E_{N}  + \frac{n^{2}}{2\nu}
    \left(1 - \nu U\right)\left(\zeta^{2}_{1} + \zeta^{2}_{2}+ \zeta^{2}_{3}\right)
    \label{eq:EMF-su4},
\end{equation}
with $E_{N}$ the energy of the normal state.
However, the energy is now parametrized by three polarizations
$\zeta_{1}, \zeta_{2}, \zeta_{3}$ as there are four iso-spin bands~\footnote{
    The polarizations $\zeta_{i}$ are also subject to additional constraints arising from conservation of the total density. See~\onlinecite{Raines2024a,*Raines2024b} for details on the parameterization.
}.
At $\nu U >1$, the energy is minimized by making
$\zeta^2_{1} + \zeta^2_{2} + \zeta^2_{3}$ the largest possible, hence there is again a strong first-order transition accompanied by the divergence of susceptibility at $\nu U$ approaching $1$ from below.
The resulting state
is a highly degenerate quarter-metal state, with 3 spontaneously depopulated bands and one band containing all fermions (see~\cite{Raines2024a,*Raines2024b} for details).

\begin{figure}
    \centering
    \includegraphics[width=0.8\linewidth]{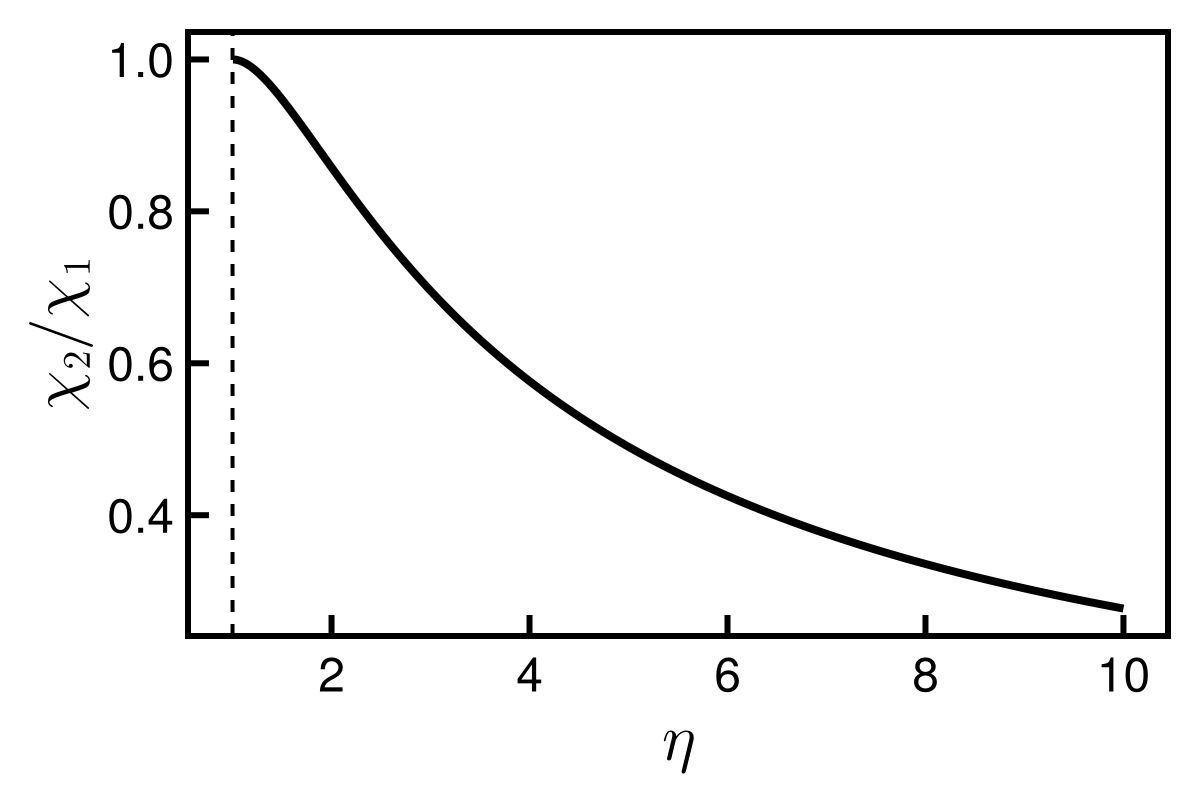}
    \caption{Suppression of the inter-valley susceptibility $\chi_{2}$ relative to the intra-valley susceptibility $\chi_{1}$ as a function of the anisotropy parameter $\eta$.\label{fig:chi_inter}}
\end{figure}

For finite anisotropy ($\eta\neq1$) the degeneracy is partly lifted, with IVC and sIVC states becoming energetically disfavored.
This can be seen by comparing the bare static susceptibilities in each channel; these now take
different values depending on whether the associated order is at small momentum (VP, FM, SFM) or large momentum (IVC, sIVC).
Concretely $\chi_{0}^{VP}=\chi_{0}^{FM}=\chi_{0}^{SFM}=\chi_{1}$, and $\chi^{IVC}_{0}=\chi^{sIVC}_{0}=\chi_{2}$ with
\begin{equation}
    \begin{gathered}
        \chi_{1}
        = - 2\lim_{\mathbf{q}\to0}\sum_{\mathbf{k}\tau} \frac{n_{F}(\epsilon_{\mathbf{k}+\mathbf{q}/2,\tau}) - n_{F}(\epsilon_{\mathbf{k}-\mathbf{q}/2,\tau})}{\epsilon_{\mathbf{k}+\mathbf{q}/2,\tau} -\epsilon_{\mathbf{k}-\mathbf{q}/2,\tau}},\\
        \chi_{2}
        = - 2\sum_{\mathbf{k}\tau} \frac{n_{F}(\epsilon_{\mathbf{k},\tau}) - n_{F}(\epsilon_{\mathbf{k},\tau})}{\epsilon_{\mathbf{k},\tau} -\epsilon_{\mathbf{k},-\tau}}.
    \end{gathered}
\end{equation}
At $T =0$ these evaluate to
\begin{equation}
    \chi_{1}
    = 4 \nu,\quad
    \chi_{2}= \chi_{1}\frac{4}{\eta - \eta^{-1}}
    \left(\tan^{-1}\eta -\frac{\pi}{4}\right).
    \label{eq:static-susc}
\end{equation}
The ratio
$\chi_{2}/\chi_1$ is a decreasing function of $\eta > 1$ (see \cref{fig:chi_inter}) and thus anisotropy favors
zero momentum states, i.e.,
VP, FM, and SFM orders.
These three, however, remain degenerate as long as $U_1 =U_2$, and there is still a
first-order transition to a quarter-metal state (three empty bands and one filled band) at $\nu U = 1$, where
VP, FM and SFM susceptibilities diverge.
The only difference with the SU(4) case is the number of Goldstone modes in the quarter-metal state.

\begin{figure*}
    \centering
    \includegraphics[width=0.9\linewidth]{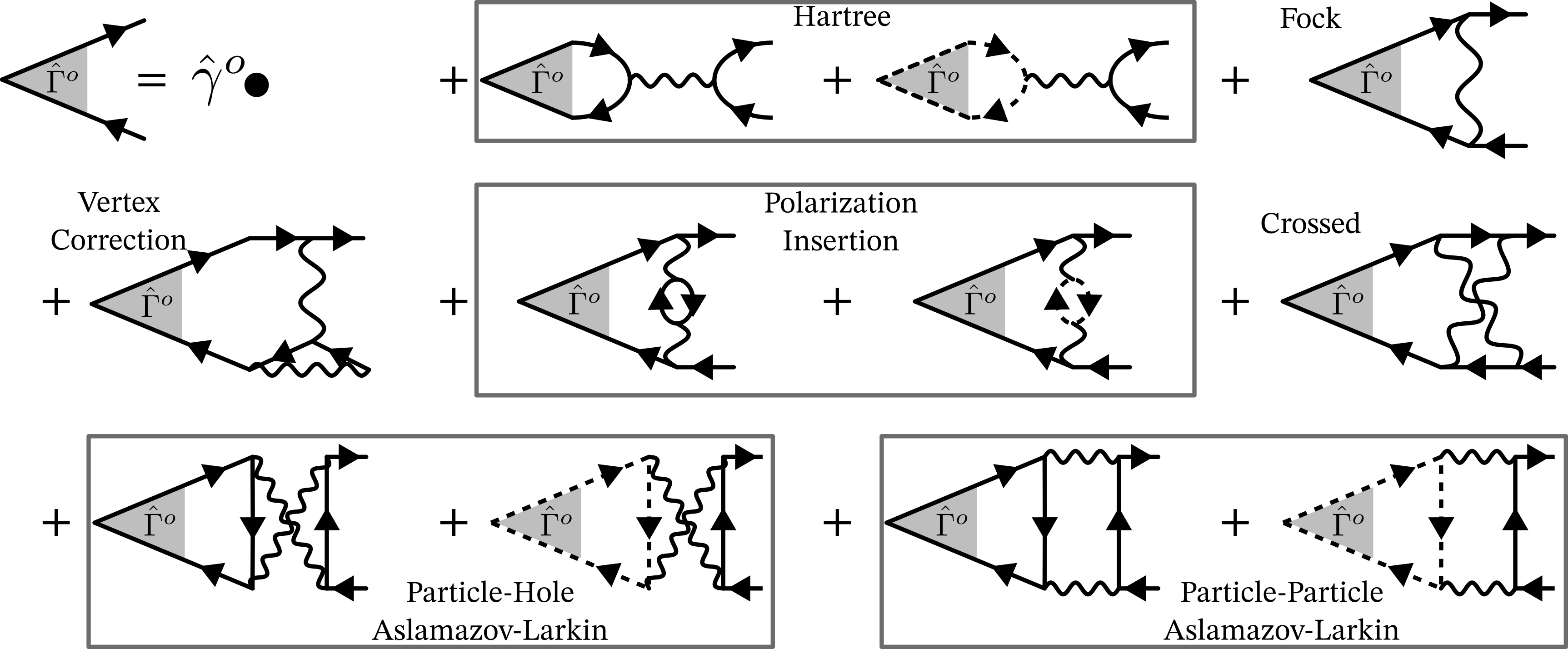}
    \caption{Contributions to the renormalized Bethe-Salpeter equation for the iso-spin vertices to two loop order.
        Solid fermionic lines are in the $\tau=+$ valley while dashed lines are in the $\tau=-$ valley.
        The diagrammatic expressions for dashed vertices are the same (not shown).
        For the SFM and VP order parameters the solid and dashed vertices have equal magnitudes and opposite signs, while for the
        FM order they are identical.
        \label{fig:iso-spin-bs}}
\end{figure*}

When $U_1 \neq U_2$, FM and SFM orders are still degenerate, but the degeneracy between VP and FM/SFM is lost.
To obtain dressed VP and FM/SFM susceptibilities we can again
calculate the renormalized vertices
\begin{equation}
    \hat{\Gamma}^{o} \equiv \Gamma^{o}\hat{\gamma}^{o},
\end{equation}
where $o=$ VP, FM/SFM, $\hat{\gamma}^\text{VP} = \hat{\tau}_{z}$, and  $\hat{\gamma}^\text{FM/SFM}= \hat{\sigma}_{z}$.
At the mean-field level the vertices obey a one-loop Bethe-Salpeter equation of the form
\begin{equation}
    \hat{\Gamma}^{o} = \hat{\gamma}^{o} +
    \hat{\Gamma}^{o}_{H} + \hat{\Gamma}^{o}_{F},
\end{equation}
where $\hat{\Gamma}^{o}_{H}$ and $\hat{\Gamma}^{o}_{F}$ denote Hartree and Fock
contributions, specified in
\cref{fig:iso-spin-bs}.
The matrix structure of the vertices is not changed by renormalization, hence $\hat{\Gamma}^{o} = \hat{\gamma}^{o} \Gamma^{o}$, where
\begin{equation}
    \Gamma^{o} \equiv \frac{1}{4}\tr[\hat{\gamma^{o}}\hat{\Gamma}^{o}]
    = 1 +
    \Gamma^{o}_{H} + \Gamma^{o}_{F},
    \label{eq:scalar-bs}
\end{equation}
and the susceptibility $\chi^{o}$ is related to $\Gamma^{o}$ by
\begin{equation}
    \chi^{o} \equiv 4\nu \Gamma^{o}.
    \label{eq:chio-gammao}
\end{equation}

The Fock diagram is identical for the VP and FM/SFM channels:
\begin{equation}
    \Gamma^{o}_{F} = -U_{1}\nu T\sum_{n',\mathbf{k}'}\frac{1}{4}\tr[(\hat{\gamma}^{o}\hat{G}_{0;k'})^{2}]\Gamma^{o}
    = \frac{1}{4}U_{1}\chi_{1}\Gamma^{o}.
    \label{eq:XF}
\end{equation}
Here $\hat{G}_{0}$ is the normal state matrix Green's function with spin $\sigma$ and valley $\tau$ indices, i.e.,
\begin{equation}     \hat{G}_{0;k,\tau\tau'\sigma\sigma'} = G_{0;k,\tau}\delta_{\tau\tau'}\delta_{\sigma\sigma'}
    \label{eq:matrix-G0}
\end{equation}
The two Hartree diagrams, on the other hand, are non-zero only for the VP channel:
\begin{equation}
    \Gamma^{VP}_{H}  = \frac{1}{2}(U_{2} - U_{1})\chi_{1}
    \Gamma^{VP},
\end{equation}
while $\Gamma^{\text{FM/SFM}}_{H}=0$.
Plugging these into the right hand side of \cref{eq:scalar-bs} we obtain the MF result for the renormalized vertices:
\begin{equation}
    \Gamma^{o}_{MF}  = \frac{1}{1 - \nu U^{o}},\quad
    U^\text{FM} = U_{1},\quad U^\text{VP} = 2U_{2} - U_{1}.
    \label{eq:1loop-4comp}
\end{equation}
For $U_1 > U_2$, FM/SFM order develops first, while for $U_2 > U_1$, VP order develops first. In each case,
there is sequence of first-order transitions: first into a half-metal state and then into a quarter metal state.
Near each transition, the corresponding susceptibility diverges.
 The mean field approximation to the Luttinger-Ward functional $\Phi$ is shown in \cref{fig:lwMF}, where now
solid fermionic lines represent the full matrix Green's functions with both spin $\sigma$ and valley $\tau$ indices,
and the wavy line represents the bare Hubbard interactions $U_{1}$ and $U_{2}$ with appropriate matrix structure in valley space (cf.\ \cref{eq:Hint}).
In analytical form,
\begin{multline}
    \Phi_{MF}[\hat{G}] =
    \frac{1}{4}T^{2}\sum_{kk', \pm}
    \left(
    - U_{1}\Tr[\hat{G}_{k}\hat{\tau}_{\pm}\hat{G}_{k'}\hat{\tau}_\pm]
    \right.\\
    +U_{1}\Tr[\hat{G}_{k}\hat{\tau}_{\pm}]\Tr[\hat{G}_{k'}\hat{\tau}_{\pm}]
    \\
    \left.+
    U_{2}\Tr[\hat{G}_{k}\hat{\tau}_{\pm}]\Tr[\hat{G}_{k'}\hat{\tau}_{\mp}]\right),
    \label{eq:lwMF} \end{multline} where the trace is now over both spin and valley and $\hat{\tau}_{\pm}=(1/2)(\hat{\tau}_{1}\pm i \hat{\tau}_{2})$ are raising (lowering) Pauli matrices in valley space.
Minimizing \cref{eq:E-LW-var} with respect to $\hat{G}$ one obtains the mean-field Green's function $\hat{G}_{MF}$ and the mean-field ground state energy~\cref{eq:EMF-su4}, from $E_{LW}[\hat{G}_{MF}]$.

\subsection{Iso-spin susceptibility beyond MF}
\label{sec:two-valley-beyond}

We now turn to the question of how the FM/SFM and VP instabilities are modified beyond the MF approximation.
As before, we start by considering
the Bethe-Salpeter equation in VP and FM/SFM channels with two-loop diagrams for the irreducible four-point function.
We will see that in the low density limit certain diagrams are
enhanced by $\ln \Lambda/k_{F}$.
The two-loop contributions to the Bethe-Salpeter equation are shown in the second and third lines of \cref{fig:iso-spin-bs}.
They include the corrections to the side vertices in the one-loop Fock diagram, the term with inserted particle-hole polarization, the diagram with crossed interaction lines, and two types of Aslamazov-Larkin (AL) diagrams with parallel and crossed interaction lines.
We verified that logarithmic renormalizations come from the diagrams with crossed interaction lines, which contain particle-particle polarization bubble with unconstrained momentum integration.
For definiteness, we call the last diagram in the
second line in~\cref{fig:iso-spin-bs} the crossed diagram and the last two in the third line AL particle-particle diagrams.
In other diagrams, all intermediate fermions are constrained to be near the Fermi surface.
As a consequence, to logarithmic accuracy, we need only keep the crossed diagram and the particle-particle AL diagrams.
The calculation of the crossed diagram proceeds as for the spin-only case, \cref{eq:gamma-crossed-spin-only}, with the only difference being a factor of $2$ due to the extra degrees of freedom in the spin and valley model.
This diagram does not distinguish between VP and FM/SFM vertices and gives
\begin{equation}
    \Gamma^{o}_{\text{crossed}} = U_{1}^{2} T^{2}
    \sum_{k',q} G^{2}_{0;k'\tau}G_{0;k-q,\tau}G_{0,k'+q,\tau}
    \Gamma^{o}
\end{equation}
where $k = (\mathbf{k}, \epsilon_n)$ is the external momentum/frequency, and $|\mathbf{k}| \approx k_F$.
There is no summation over $\tau$ and the result is the same for $\tau =+$ and $\tau=-$.
One can easily verify that integration over $k'$ is constrained to the Fermi surface, but integration over $q$ is not constrained, and that the logarithmic contribution comes from  $q \gg k_{F}$.
For these $q$, $\Gamma^{o}_{\text{crossed}}$ can be approximated by
\begin{multline}
    \Gamma^{o}_{\text{crossed}} \approx U_{1}^{2}
    T \sum_{k'} G^{2}_{0;k',\tau} T\sum_q G_{0;-q,\tau} G_{0;q,\tau} \Gamma^{o}\\
    \approx -
    (U_{1}\nu)^2 \ln\frac{\Lambda}{k_{F}}
    \Gamma^{o}.
    \label{eq:gamma-crossed}
\end{multline}
where we used $-T \sum_{k} G^{2}_{0;k,\tau} = \nu$.
Notice that the anisotropy $\eta$ does not enter as it does not affect the density of states in each valley.
In contrast, the AL particle-particle diagrams distinguish between VP and FM/SFM vertices.
For the latter, these diagrams vanish due to the spin summation, i.e., \cref{eq:gamma-crossed} is the only contribution at the two-loop order.
Solving the BS equation in the FM/SFM channels we obtain, at the two-loop level,
\begin{equation}
    \Gamma^{\text{FM/SFM}}_{2\text{loop}} = \frac{1}{1 - \nu U + (\nu U)^{2}\ln{\frac{\Lambda}{k_{F}}}}.
    \label{eq:fm-sfm}
\end{equation}
Clearly, the two-loop crossed diagram suppresses the tendency towards a Stoner instability in the FM and SFM channels.

For the VP vertex,
the particle-particle AL term does not vanish and is given by
\begin{multline}
    \Gamma^{\text{VP}}_{AL;pp} = - 2 T^{2}\sum_{k',q}\left(
    U^{2}_{1}G_{0;k-q,+}G_{0;k'+q,+}G^{2}_{0;k',+}\right.\\
    \left.- U^{2}_{2}G_{0;k-q,-}G_{0;k'+q,+}G^{2}_{0;k',+}
    \right)
    \Gamma^{\text{VP}},
    \label{eq:XALpp-simp}
\end{multline}
where we have performed a $C_{4}$ rotation in the second term to simplify the expression.
The logarithmic contribution again comes from internal momenta $q \gg k_F$, when
the integration in \cref{eq:XALpp-simp} factorizes to
\begin{multline}
    \Gamma^{\text{VP}}_{AL;pp} \approx - T\sum_{k'}G^{2}_{0;k',+}\\
    \times 2T \sum_{q}\left(
    U^{2}_{1}G_{0;-q,+}G_{0;q,+}
    - U^{2}_{2}G_{0;-q,-}G_{0;q,+}
    \right) \Gamma^{\text{VP}} \\
    = \nu \left(U^{2}_{1}\Pi_{1}^{pp}-U^{2}_{2}\Pi_{2}^{pp}\right)\Gamma^{\text{VP}}
\end{multline}
where
we have defined
\begin{equation}
    \begin{gathered}
        \Pi_{1}^{pp} \equiv 2T\sum_{q} G_{0;-q,+}G_{0;q, +},\\
        \Pi_{2}^{pp} \equiv 2T\sum_{q} G_{0;-q,-} G_{0;q,+}.
    \end{gathered}
\end{equation}
For the former, we have, similar to the crossed diagram,
\begin{equation}
    \Pi^{pp}_{1}= 2\nu
    \ln\frac{\Lambda}{k_{F}}.
\end{equation}
On the other hand, $\Pi^{pp}_{2}$ involves particles at both valleys and thus depends on the anisotropy.
Evaluating this contribution we find
(see~\cref{sec:pp-bubble})
\begin{equation}
    \Pi_{2}^{pp}
    = \frac{4\eta}{\eta^{2} + 1}\ln \frac{\Lambda}{k_{F}},
\end{equation}
Putting together crossed and AL contributions, we obtain
\begin{equation}
    \Gamma^{\text{VP}}_{\text{AL}}  + \Gamma^{\text{VP}}_{\text{crossed}}
    = \left((U_1 \nu)^2 - \frac{4\eta}{\eta^{2} + 1} (U_2 \nu)^2  \right) \ln \frac{\Lambda}{k_{F}}\Gamma^{\text{VP}}.
\end{equation}
Adding the one-loop term we obtain
the two-loop expression for the VP vertex
\begin{multline}
    \Gamma^{\text{VP}}_{2\text{loop}} = \left[1  - 2 U_2 \nu \left(1 - (U_2 \nu) \frac{2\eta}{\eta^{2} + 1} \ln{\frac{\Lambda}{k_{F}}}\right) \right.\\
        \left.+ U_1 \nu\left(1 - (U_1 \nu) \ln{\frac{\Lambda}{k_{F}}}\right)\right]^{-1}
    \label{eq:vp-vertex-2loop}
\end{multline}
Comparing \cref{eq:fm-sfm,eq:vp-vertex-2loop}, we see that the two-loop renormalizations of the VP and FM/SFM vertices are different even if we set $U_1 =U_2 =U$.
In this last case,
\begin{equation}
    \begin{gathered}
        \Gamma^{\text{FM/SFM}}_{2\text{loop}} = \frac{1}{1  -  U\nu + (U\nu)^2 \ln{\frac{\Lambda}{k_{F}}}} \\
        \Gamma^{\text{VP}}_{2\text{loop}} = \frac{1}{1  -  U\nu + (U\nu)^2 \left(\frac{4\eta}{\eta^{2} + 1}-1\right)
            \ln{\frac{\Lambda}{k_{F}}}}.
    \end{gathered}
    \label{eq:U1eqU2-vertices}
\end{equation}
Notice that for $\eta >1$, the logarithmic suppression is weaker in the VP channel than in the FM/SFM channels.
Furthermore, for $\eta >2 + \sqrt{3}$, the two-loop renormalization of  $\Gamma^{\text{VP}}$ actually increases the tendency towards the Stoner instability.

As we did for the spin-only case,
we next go beyond two-loop order.
To simplify the presentation, we set $U_1 = U_2 =U$.
As before, note that at $l$-loop order, maximally crossed irreducible diagrams come with a factor of $\ln(\Lambda/k_{F})^{l-1}$ due to $l-1$ factors of $\Pi^{pp}$, while any other diagrams have lower powers of $\ln\Lambda/k_{F}$.
Evaluating the irreducible vertices with logarithmic accuracy by keeping the diagrams with the largest power of $\ln(\Lambda/k_{F})$ at each loop order, and solving then Bethe-Salpeter equations for $ \Gamma^{o}$, we obtain
\begin{equation}
    \Gamma^{FM/SFM} = \frac{1}{1- \tilde{U}_1\nu}, \quad \Gamma^{VP} = \frac{1}{1- (2\tilde{U}_2-{\tilde U}_1) \nu},
    \label{eq:vertices-ladders}
\end{equation}
which formally look the same as in MF approximation, but with the effective interactions
\begin{equation}
    \begin{gathered}
        \tilde{U}_{1} = \frac{U}{1 + \nu U\ln\frac{\Lambda}{\nu_{F}}},\\
        \tilde{U}_{2} = \frac{U}{1 + \nu U\frac{2\eta}{\eta^{2}+1}\ln\frac{\Lambda}{k_{F}}}.
    \end{gathered}
\end{equation}
The corresponding susceptibilities are
\begin{equation}
    \chi^{\text{FM}/\text{SFM}} = \frac{4\nu}{1- \tilde{U}_1\nu}, \quad \chi^{\text{VP}} = \frac{4\nu}{1- (2\tilde{U}_2-{\tilde U}_1) \nu},
    \label{eq:susc-ladders}
\end{equation}

\begin{figure}
    \centering
    \includegraphics[width=\linewidth]{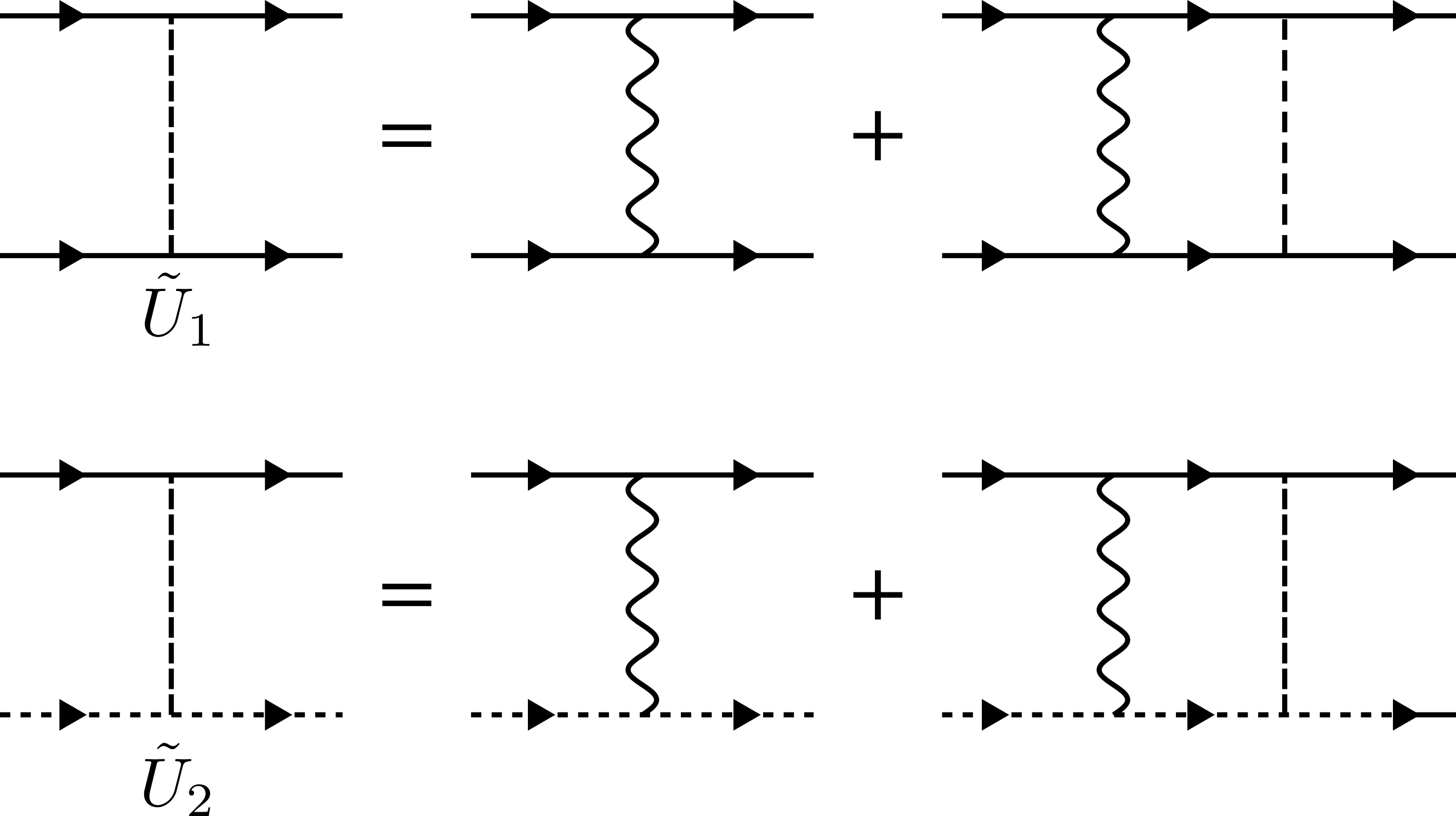}
    \caption{Leading renormalization of the intra-valley density-density interaction $U_1$ and inter-valley density-density interaction $U_2$}
    \label{fig:renormalized}
\end{figure}

In Ref.~[\onlinecite{Calvera2024}] the same $\tilde{U}_{1}$ and $\tilde{U}_{2}$ have been obtained
by summing ladder series of particle-particle diagrams for the renormalized interactions (see \cref{fig:renormalized}).

\begin{figure}
    \centering
    \includegraphics[width=0.8\linewidth]{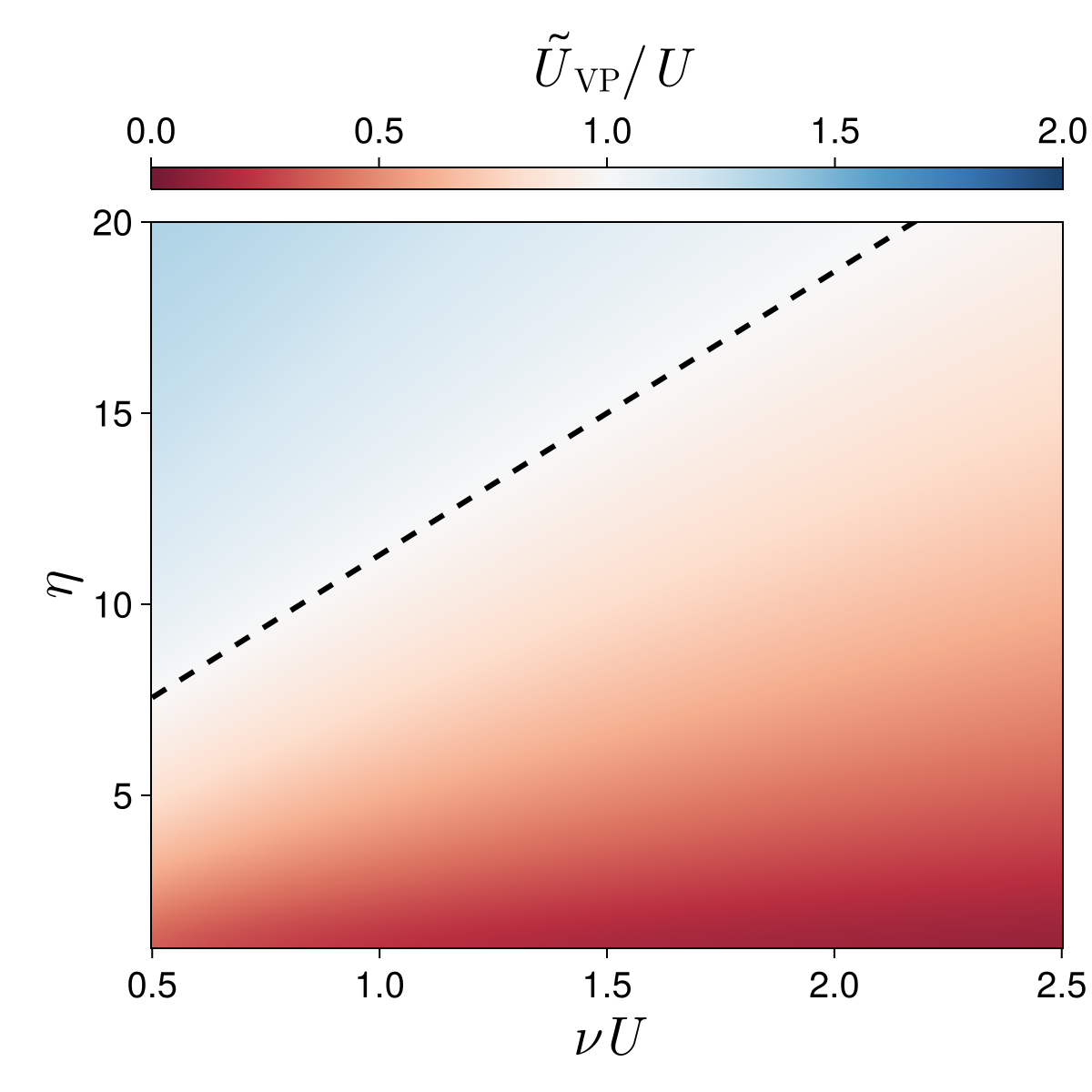}
    \caption{
        (Color online)
        Renormalized value of the valley-polarization coupling $\tilde{U}_{VP}$ compared to its bare value $U_{VP}=U$,
        as a function of the bare coupling $U$ and anisotropy parameter $\eta$.
        The dashed line indicates where $\tilde{U}_{VP}=U$.
        Notably, for sufficiently strong anisotropy, $\tilde{U}_{VP} > U$.
        \label{fig:renormalized_VP}
    }
\end{figure}

\subsection{Ground state energy beyond MF}
\label{sec:gs}

To obtain the ground state, we employ the Luttinger-Ward functional as we did for the spin-only case, \cref{sec:gs-spin}.
Specifically we sum the logarithmically enhanced terms in \cref{fig:lw-ladder}, where
solid fermionic lines now represent matrix Green's functions with both spin $\sigma$ and valley $\tau$ indices, and the interaction lines include both intra- and inter-valley interactions $U_{1}$ and $U_{2}$.
Recall that the MF approximation~\cref{fig:lwMF} corresponds to keeping the first diagram in each of the two lines in \cref{fig:lw-ladder}.

Again, the leading contribution to each diagram comes from when all but two pairs of particles are away from the Fermi surface, and the remaining particles are near the Fermi surface, \cref{fig:lw-effective-MF},
so the combinatoric factors associated with ways of choosing $2$ out of $2l$ lines to be near the Fermi surface in an $l-$loop diagram cancel the overall factor of $1/l$.
Therefore, like in the spin-only case,
the Luttinger-Ward functional
\emph{retains the same form as in MF}, \cref{eq:lwMF},
but with the
effective interactions $\tilde{U}_{1}$ and $\tilde{U}_{2}$.
In explicit form,
\begin{multline}
    \Phi_{\log}[\hat{G}]\approx
    \frac{1}{4}T^{2}\sum_{kk', \pm}
    \left(
    - \tilde{U}_{1}\Tr[\hat{G}_{k}\hat{\tau}_{\pm}\hat{G}_{k'}\hat{\tau}_\pm]
    \right.\\
    +\tilde{U}_{1}\Tr[\hat{G}_{k}\hat{\tau}_{\pm}]\Tr[\hat{G}_{k'}\hat{\tau}_{\pm}]
    \\
    \left.+
    \tilde{U}_{2}\Tr[\hat{G}_{k}\hat{\tau}_{\pm}]\Tr[\hat{G}_{k'}\hat{\tau}_{\mp}]\right).
    \label{eq:lw-effective-MF}
\end{multline}

The outcome of this analysis is the following:
\begin{enumerate}
    \item The MF-like description of the Stoner transition in a two-valley system remains valid once we include leading logarithmic renormalizations beyond MF, but with the original interactions $U_1$ and $U_2$ renormalized to effective interactions ${\tilde U}_1$ and ${\tilde U}_2$.
    \item For a circular Fermi surface, both ${\tilde U}_1$ and ${\tilde U}_2$ are strongly reduced compared to the original interactions such that as long as the theory remains under control, the condition for a Stoner instability in any channel is not satisfied.
          Whether the instability develops at even larger $U$ is beyond the scope of this paper.
    \item For an elliptical Fermi surface in each valley, governed by $\eta >1$, ${\tilde U}_1$ is still strongly reduced, but ${\tilde U}_2$ is reduced less.
          In the limit of strong anisotropy, $\tilde{U}_{2}\approx U_{2}$.
    \item As a consequence,  Stoner transitions into FM and SFM states do not occur within the logarithmic approximation, but a Stoner transition into the VP state does occur if the anisotropy is strong enough.
    \item For $\eta \gg 1$,  the Stoner instability towards VP occurs at smaller $U$ than at the mean field level ($\tilde{U}^{VP} > U^{VP}$).
    \item The VP susceptibility diverges upon approaching the Stoner instability from below, and is zero on the other side of the transition where the VP order parameter jumps to its maximum value.
          The same happens near a FM/SFM transition if we treat ${\tilde U}_1$ as a parameter which can reach $\nu {\tilde U}_1 =1$
\end{enumerate}
This last item is the most remarkable outcome of the current treatment: the unconventional nature of Stoner transitions in 2D is preserved in the logarithmic approximation beyond MF.\@
Namely, the transitions are strongly first order into maximally ordered states, yet the corresponding susceptibility diverges as the system approaches an instability from the disordered side.

\begin{figure}
    \centering
    \vspace{1em}
    \includegraphics[width=\linewidth]{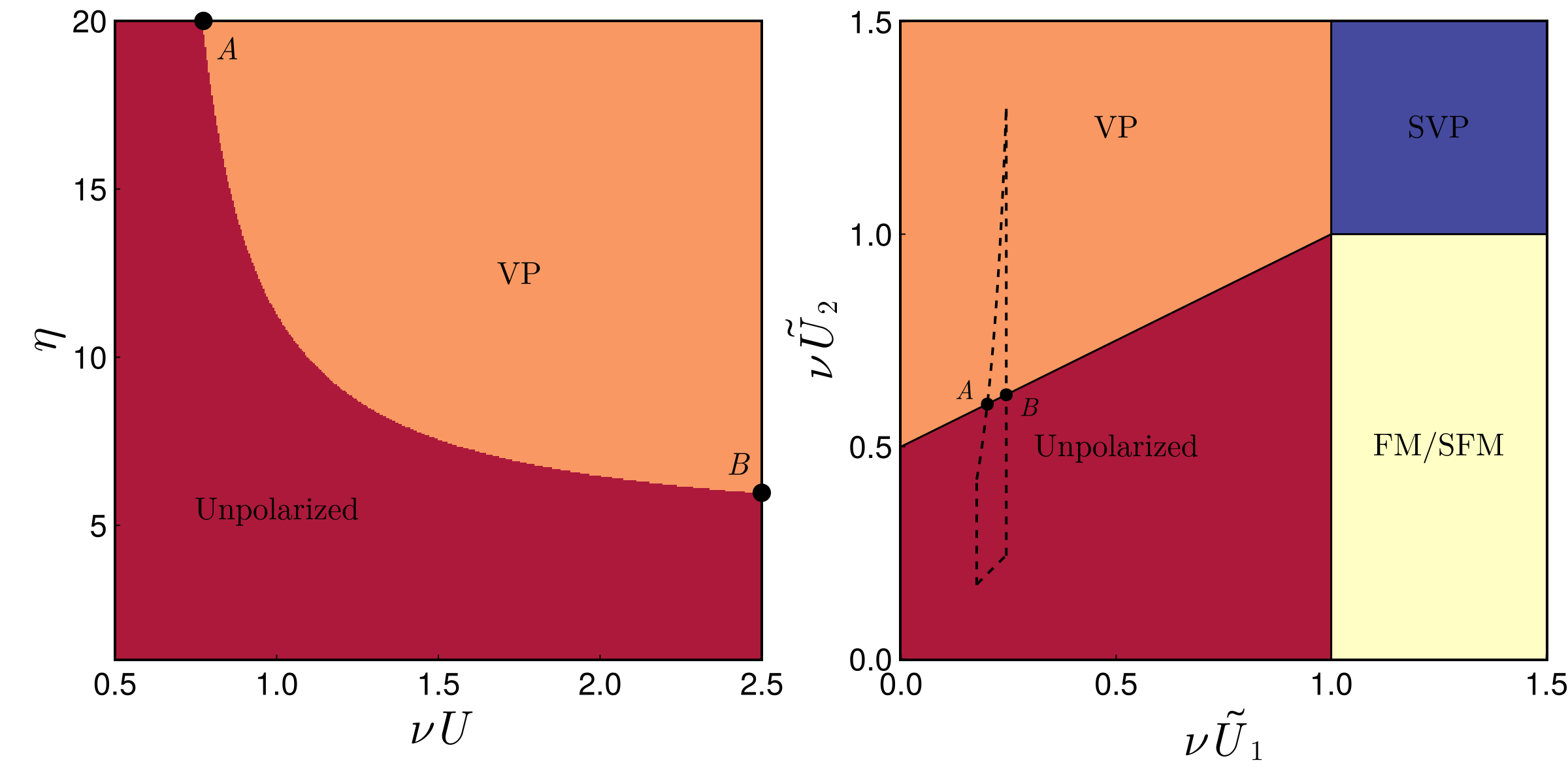}
    \caption{(Color online) Left panel: Phase diagram within the logaritmic approximation as a function of bare coupling $U_{1}=U_{2}=U$ and anisotropy parameter $\eta$ for $\Lambda/k_{F}=40$.
    For small anisotropy the system remains in the unpolarized state.
    For larger $\eta$, there is a first order transition into a fully VP state.
    The VP susceptibility diverges as the transition is approached from the unpolarized region.
    Right panel: The phase diagram in terms of renormalized interactions $\tilde{U}_{1}$ and $\tilde{U}_{2}$, treated as parameters.
    This phase diagram is the same as within MF.
    Within our logarithmic approximation,  the FM/SFM region and
    the region of spin and valley polarization (SVP)
    are inaccessible.
    The region corresponding to the left panel is delineated by the dashed lines.
    The points labeled $A$ and $B$ into the left panel map onto the correspondingly labeled points in the right panel.
    \label{fig:aniso-phase}}
\end{figure}

In \cref{fig:renormalized_VP} we plot the effective ${\tilde U}^{VP} = 2 {\tilde U}_2 - {\tilde U}_1$ vs the bare
interaction in the VP channel, $2 U_2 - U_1$, which is $U$ when $U_1=U_2 = U$.
At large $\eta$, the condition
${\tilde U}_{VP} \nu =1$ is reached at smaller $U$ than the condition $U \nu =1$, i.e., the Stoner transition into the VP state occurs at smaller $U$ than in MF.
In  \cref{fig:aniso-phase} we present the phase diagrams.
The left panel is the phase diagram obtained within our logarithmic approximation.
A FM/SFM order does not develop, but VP order develops and the critical $U$ get reduced as $\eta$ increases.
The right panel is the formal phase diagram obtained by treating ${\tilde U}_1$ and ${\tilde U}_2$ as variables.
The region inside the dashed lines is the one in the left panel.

\section{Comparison with RPA}
\label{sec:rpa}

It is worth comparing our conclusions with those obtained in Ref.~[\onlinecite{Calvera2024}] for the Coulomb interaction treated within the RPA.\@
The RPA approximation to the energy
employed therein
amounts to
summing all ring diagrams of particle-hole bubbles.
We, on the contrary,
estimated the Luttinger-Ward functional as a series of particle-particle bubbles in the full Green's function, corresponding to keeping only the largest contribution at each loop-order.
There are some differences in these approaches due to directly using the diagrammatic expansion of the energy vs.\ the Luttinger-Ward formalism,
which we cover in~\cref{sec:self-consistent}, but we first discuss the primary differences which arise due to the summation of particle-particle vs.\ particle-hole bubbles and the implications for experiments in AlAs.

Minimizing the RPA ground state energy, the authors of~[\onlinecite{Calvera2024}] obtained the phase diagram as function of $\eta$ and $r_s$.
By increasing $r_s$ at large enough $\eta$, they found a first-order transition into the VP state and then another first-order transition into spin and valley-polarized state (SVP).
The critical $r_s$ for the VP decreases as $\eta$ increases.
These results are in line with earlier variational Quantum Monte Carlo calculation by the same group~\cite{Valenti2023}.

The sequence of ordered states and the reduction of the critical interaction for VP for larger $\eta$ are consistent with our results.
However, the VP and FM/SFM susceptibilities obtained in~[\onlinecite{Calvera2024}] do not display a Stoner-like divergence. The difference comes from the fact that the RPA is an infinite series in particle-hole bubbles, while in our calculation we included an infinite series in particle-particle bubbles.
At second order, both approaches include the second diagram in~\cref{fig:lw-ladder},
which introduces a difference between VP and FM/SFM order, and hence we all found that the tendency towards VP is stronger.
But higher-loop diagrams are different in their and our approaches.
Unlike our case, there is no cancellation of the diagrammatic combinatoric factors $1/l$ in the RPA,
and hence there is no particular relationship between the condition for the first-order transition and the behavior of the susceptibility at the transition point.

That the susceptibilities do not diverge within RPA can be qualitatively understood by noticing that the RPA screened Coulomb interaction is
\begin{equation}
    \nu U_{\text{RPA}}(q) = \frac{r_{s} k_{F}}{q+4r_{s}k_{F}} \leq \frac{1}{4},
\end{equation}
where $r_{s}k_{F} = 4\pi e^{2}\nu$ and the factor of $4$ comes from the spin and valley sums.
The screened Coulomb interaction never approaches the point $\nu U_{\text{RPA}}(q) =1$, where the susceptibility would diverge.

On physical grounds, it is crucial to our considerations that the cutoff scale of the interaction $\Lambda$ be large compared to the Fermi momentum.
For the Hubbard-like model we consider $\Lambda$ is of order the Brillouin zone size.
In this situation, if $k_F$ is far smaller than $\Lambda$, our treatment involves keeping
only
the largest terms $\propto [\ln(\Lambda/k_{F})]^{n}\gg1$
at each order, while the terms included in the RPA are of order $1$, i.e., are smaller.
However, for systems where the interaction is more long-ranged, the cutoff scale $\Lambda$ could be instead related to a property of the interaction itself.
In particular, the authors of~[\onlinecite{Calvera2024}] argued that if $\Lambda$ is related to the screening gate separation, it may be comparable to $k_{F}$, and $\ln \Lambda/k_{F}$ is no longer a large parameter.
In this situation,
particle-hole diagrams can no longer be neglected, but contributions from the particle-particle channel cannot be neglected either.

\subsection{Self-consistency: Conserving vs. Non-conserving}
\label{sec:self-consistent}

One other difference between the approach employed in ~[\onlinecite{Calvera2024}] and this work relates to self-consistent nature of the Luttinger-Ward calculation, and correspondingly the lack of self-consistency in the RPA calculation.
Note, that in general including the particle-hole ring diagrams in the Luttinger-Ward functional is different from
evaluating the diagrams of the same type in
the RPA approach
(Ref.~[\onlinecite{Calvera2024}]).
Within the Luttinger-Ward formalism the Green's function is to be calculated fully self-consistently.
The RPA summation on the other hand uses a free-fermion-like ansatz for the Green's function with an iso-spin dependent chemical potential.

As noted in \cref{sec:gs-mf-spin},
for the Hubbard model the distinction does not matter
for
either the MF or logarithmic approximations,
as the self-energy obtained from the Luttinger-Ward functional is constant.
Thus, the Green's function is of free-fermion-like form and it does not matter where we use the Luttinger-Ward functional or calculate the energy directly from the variational Green's function.

On the other hand, if one considers the Luttinger-Ward functional obtained from summing ring diagrams, sometimes called the self-consistent GW approximation,
the self-energy
is in general a function of frequency and momentum.
The RPA summation scheme thus differs in that the self-energy equation is not evaluated self-consistently (this is sometimes called the $G_0W$ approximation).
For this reason, the RPA
is in principle not a
fully self-consistent conserving approximation
as opposed to calculations within the Luttinger-Ward functional.
Of course, self-consistency does not guarantee the best approximation.

\subsection{Application to AlAs}

The phase diagram of AlAs quantum wells as a function of electron density has been thoroughly investigated in recent years~\cite{Gunawan2006,Shayegan2006, Hossain2020,Hossain2021,Hossain2022}.
As mentioned in \cref{sec:two-valley-mf}
bulk AlAs has ellipsoidal Fermi pockets near the three inequivalent points in the Brillouin zone.
When confined along one of the periodic axes, e.g.
$[001]$, and placed atop a GaAs substrate, the energy spectrum splits between lower-energy excitations in the two in-plane $X$ and $Y$ valleys and
higher-energy excitations in the out-of-plane $Z$ valley~\cite{Shayegan2006}.
When the Fermi level is placed in $X/Y$ valleys, the low-energy excitations are located in the $X$ and $Y$ valleys, near elliptical Fermi pockets centered at $X =[100]$, and $Y =[010]$ and related by a $C_{4}$ rotation which exchanges the $X$ and $Y$ points.

Resistivity measurements in the presence of an external strain and magnetic field reveal two ordered states that appear
upon decreasing electron density, i.e., upon increasing $r_s$.
At the first critical $r_{s,1}$,  the system undergoes a strong first-order transition into a VP state, at which one of the valleys moves above the Fermi level (a half-metal state).
Then, at $r_{s,2} > r_{s,1}$, there is another first-order transition into SVP state, when the spins in the remaining valley order ferromagnetically and the band with one spin projection moves above the Fermi level (a quarter-metal state).
The experimentally observed sequence of transitions into VP and then SVP states has been reproduced in variational Monte Carlo calculations for a two-valley system with elliptical Fermi surfaces and
Coulomb interaction~\cite{Valenti2023}, although at smaller $r_s$ than in the experiments.
These studies also found that the critical $r_s$ for the VP transition gets smaller with increasing ellipticity of the Fermi surfaces, i.e., with increasing $\eta$.
A
similar phase diagram holds also for the Hubbard interaction~\cite{Berg2024}

The sequence of first-order transitions, the disappearance of some of Fermi surfaces immediately above the transition, and the reduction of the threshold for VP with increasing $\eta$  is consistent with our results and with the RPA calculations in Ref.~[\onlinecite{Calvera2024}].
The distinguishing feature of our theory is the prediction that the transitions remain Stoner-like, i.e., that the VP and spin susceptibilities diverge upon approaching the corresponding transitions from a disordered side.
We emphasize that this holds in the logarithmic approximation that we used.
Beyond this approximation, when RPA-type corrections to the Luttinger-Ward functional (ring diagrams) are included, the susceptibility saturates at a finite value at each transition.
However, if $\ln{\Lambda/k_F}$ is sizable, our theory predicts that the susceptibilities should still be substantially enhanced near the VP and FM transitions.

In Refs.~[\onlinecite{Gunawan2006,Shayegan2006,Hossain2021}], the authors
where able
to measure the valley susceptibility in AlAs
with high precision.
The experiments measured the valley occupation as a function of strain, which acted as a valley dependent potential, using two complementary methods.
One method relies on the anisotropic nature of the $X$ and $Y$ Fermi pockets, allowing the resistivities $R_{[100]},R_{[010]}$, along the $[100]$ and $[010]$ directions, respectively, act as a direct proxy for the densities in the corresponding valleys.
The other obtained the valley densities from the characteristic frequencies of Shubnikov-de Haas oscillations
In either case, the VP susceptibility can then be obtained as, $\chi^{VP} \propto d(n_{X} - n_{Y})/d\epsilon$.
Both methods found, with good agreement, that the valley susceptibility grows as the density decreases, corresponding to larger $r_{s}$, but the susceptibility in the immediate vicinity of critical $r_{s}$ was not measured.

Additionally, the
authors of~[\onlinecite{Hossain2020}]
extracted the spin susceptibility from their experimental measurements of magnetoresistance near
the subsequent spin transition and argued that it strongly (by more than an order of magnitude) increases
as the transition is approached.
While this regime is inaccessible within our approach, this behavior is in line with the prediction that within logarithmic approximation both transitions remain Stoner-like with the same behavior as in MF.\@

\section{Discussion and Conclusion}
\label{sec:conclusion}

In this paper, we analyzed the iso-spin instabilities of a low-density 2D Fermi gas beyond MF theory.
We showed that particle-particle ladder renormalization of the interactions are parametrically the largest contribution in the large parameter $\ln \Lambda/k_{F}$.
For an isotropic system this leads to a suppression of the interaction strength such that $\tilde{U}\nu < 1$ and Stoner instabilities cannot occur within the regime of applicability of this approximation.
It should, however, be emphasized that this does not preclude the occurrence of itinerant ferromagnet --- indeed it has been proven that such states occur on e.g., the Lieb lattice~\cite{Lieb1989} --- but rather that the instability must occur outside the regime of applicability of the logarithmic approximation.

When anisotropy is introduced in the form of a valley-dependent Fermi surface distortion, we showed that the renormalizations uniquely favor a valley polarization order, which for large enough anisotropy can develop within the logarithmic approximation.
Interestingly, the unconventional features of the MF instability persist --- namely, the VP transition is first-order into a half-metal state, with one valley completely depopulated, and the valley iso-spin susceptibility diverges as the transition point is approached from the unpolarized state.
Thus,
a first-order Stoner transition is not
an artifact of MF theory and, at least in some parameter regimes, remains when beyond MF corrections are included.
A first order transition into a VP state with a
growing VP susceptibility has been observed experimentally in AlAs~\cite{Hossain2021}.

\begin{acknowledgments}
We thank A.A. Allocca, C. Batista, E. Berg, L. Glazman, S. Huber, S. Kivelson,  L. Levitov, D. Maslov, J.H. Wilson
and particularly M. Randeria
for fruitful discussions and feedback.
The work was supported by the U.S. Department of Energy, Office of Science, Basic Energy Sciences, under Award No.
DE-SC0014402 and by the Fine Theoretical Physics Institute at the University of Minnesota.
\end{acknowledgments}

\bibliography{references.gen}

\appendix

\section{Intra- and Inter-band susceptibilities} \label{sec:susc}

In this section, we present the details of the calculation of the susceptibilities in \cref{eq:static-susc} at $T\to0$.

For the intra-band susceptibility we have
\begin{multline}
    \chi_{1} = - 2\lim_{\mathbf{q}\to0}\sum_{\mathbf{k}\tau} \frac{\Theta(E_{F}- \epsilon_{\mathbf{k}+\mathbf{q}/2,\tau}) - \Theta(E_{F} -\epsilon_{\mathbf{k}-\mathbf{q}/2,\tau})}{\epsilon_{\mathbf{k}+\mathbf{q}/2,\tau} -\epsilon_{\mathbf{k}-\mathbf{q}/2,\tau}},\\
    = 2\sum_{\mathbf{k}\tau} \delta(\epsilon_{\mathbf{k}\tau})
    = 4 \nu_{F}.
\end{multline}

The inter-band case requires more work.
First taking the $T\to 0$ limit we have
\begin{equation}
    \chi_2 = - 2\sum_{\mathbf{k}\tau} \frac{\Theta(E_{F}-\epsilon_{\mathbf{k},\tau}) - \Theta(E_{F}-\epsilon_{\mathbf{k},\tau})}{\epsilon_{\mathbf{k},\tau} -\epsilon_{\mathbf{k},-\tau}}.
\end{equation}
Defining $\epsilon = k^{2}/2m$ and $f_{\tau}(\theta)= \eta^{\tau}\cos^{2}\theta+\eta^{-\tau}\sin^{2}\theta$, and dividing the angular integration into identical regions
we can write
\begin{multline}
    \chi_2
    = 4 \nu_{F} \frac{4}{\pi}\int_{0}^{\pi/4} d\theta \int^{E_{F}/f_{-}(\theta)}_{E_{F}/f_{+}(\theta)}d\epsilon\frac{1}{\epsilon(f_{+}(\theta)- f_{-}(\theta))}\\
    = 4\nu_{F}\frac{4}{\pi}\int_{0}^{\pi/4} d\theta \frac{1}{f_{+}(\theta)-f_{-}(\theta)}\ln\frac{f_{+}(\theta)}{f_{-}(\theta)}\\
    = 4\nu_{F}\frac{4}{\pi(\eta - \eta^{-1})}\int_{0}^{\pi/4} d\theta\sec^{2}(\theta)\\
    \times \frac{1}{1 - \tan^{2}\theta}\ln\frac{\eta + \eta^{-1}\tan^{2}\theta}{\eta^{-1} + \eta\tan^{2}\theta}\\
    = 4\nu_{F}\frac{4}{\pi(\eta - \eta^{-1})}\int_{0}^{1} dz \frac{1}{1 -z^{2}}\ln\frac{\eta^{2} + z^{2}}{1+ \eta^{2} z^{2}}.
\end{multline}
Taking the derivative of the integral with respect to $\eta$
\begin{multline}
    \frac{\partial}{\partial \eta}\int_{0}^{1} dz \frac{1}{1 -z^{2}}\ln\frac{\eta^{2} + z^{2}}{1+ \eta^{2} z^{2}}\\
    = \int_{0}^{1} dz \frac{1}{1 -z^{2}}\left(
    \frac{2\eta}{\eta^{2}+ z^{2}}
    - \frac{2\eta z^{2}}{1 + \eta^{2}z^{2}}
    \right)\\
    = \int_{0}^{1} dz \frac{2\eta}{1 +\eta^{2}}\left(
    \frac{1}{\eta^{2}+ z^{2}}
    + \frac{1}{1 + \eta^{2}z^{2}}
    \right)\\
    = \frac{2\eta}{1 +\eta^{2}}\int_{0}^{\infty} dz
    \frac{1}{\eta^{2}+ z^{2}}
\end{multline}
where in the last line we made the change of variables $z\to1/z$ in the second term.
The integration is now elementary
\begin{multline}
    \chi_2 =  4\nu_{F}\frac{4}{\pi(\eta - \eta^{-1})}
    \int_{1}^{\eta} d\eta'
    \frac{2\eta}{1 +(\eta')^{2}}\int_{0}^{\infty} dz
    \frac{1}{(\eta')^{2}+ z^{2}}\\
    = 4\nu_{F}\frac{4}{\eta - \eta^{-1}}
    \int_{1}^{\eta} d\eta'
    \frac{1}{1 +(\eta')^{2}}\\
    = 4\nu_{F}\frac{4}{\eta - \eta^{-1}}
    \left(\tan^{-1}\eta -\frac{\pi}{4}\right).
\end{multline}

\section{Intra- and Inter-valley particle-particle bubbles}
\label{sec:pp-bubble}

The evaluation of the particle-particle bubbles for the two valley case is qualitatively similar to the spin-only case.
We wish to evaluate the intra-valley bubble $\Pi_{1}$ and inter-valley bubble $\Pi_{2}$:
\begin{equation}
    \begin{gathered}
        \Pi_{1}^{pp} \equiv 2T\sum_{q} G_{-q,+}G_{q, +}, \\
        \Pi_{2}^{pp} \equiv 2T\sum_{q} G_{-q,-} G_{q,+}.
    \end{gathered}
\end{equation}
For the former, we have, similar to the crossed diagram,
\begin{equation}
    \Pi^{pp}_{1}=
    \int \frac{d\omega}{2\pi} \int^{\Lambda^{2}/2m}_{k^{2}_{F}/2m}\frac{d\xi}{\omega^{2}+\xi^{2}}
    \approx
    2\nu
    \ln\frac{\Lambda}{k_{F}}
\end{equation}
where the extra factor of $2$ comes from the spin degree of a freedom.
Notice that the anisotropy $\eta$ does not enter as it does not affect the density of states.
On the other hand, $\Pi^{pp}_{2}$ involves particles at both valleys and thus depends on the anisotropy.
Defining $\epsilon = k^{2}/2m$
\begin{multline}
    \Pi_{2}^{pp}
    = 2\int \frac{d\omega}{2\pi} \int^{\Lambda^{2}/2m}_{k^{2}_{F}/2m} d\epsilon \oint \frac{d\theta}{2\pi}\\
    \times
    \frac{1}{\omega + iE_{F} - i\epsilon(\eta \cos^{2}\theta + \eta^{-1} \sin^{2}\theta)}\\
    \times
    \frac{1}{\omega - iE_{F} + i\epsilon(\eta^{-1} \cos^{2}\theta + \eta\sin^{2}\theta)}
\end{multline}
We can neglect the factor of $2E_{F}$ in the denominator since the energy of the relevant electrons is much larger,
and we obtain
\begin{equation}
    \Pi_{2}^{pp}
    \approx \frac{2}{\eta^{-1} + \eta}\int^{\Lambda^{2}}_{k_{F}^{2}}
    \frac{dx}{x}
    = \frac{4\eta}{\eta^{2} + 1}\ln \frac{\Lambda}{k_{F}},
\end{equation}

\section{Leading log contribution from dual gate screened Coulomb}

The dual gate screened Coulomb potential has the form
\begin{equation}
    \nu U(q) = r_{s} k_{F} \frac{\tanh qd}{q}
\end{equation}
where $d$ is the separation between the plates.
Working with such an interaction instead of a constant interaction the relevant particle-particle bubble is e.g.,
\begin{multline}
    \Pi^{pp}_{1}\approx 2r_{s}k_{F}\int \frac{d\omega}{2\pi} \int^{\infty}_{k_{F}}\frac{qdq}{\omega^{2}+\xi_{q}^{2}}\frac{\tanh q d}{q}\\
    = r_{s}k_{F}\int^{\infty}_{k_{F}}\frac{dq}{\frac{q^{2}}{2m}-E_{F}}\tanh q d\\
    \approx r_{s} 2m k_{F} d \int^{\infty}_{k_{F}d}\frac{dx}{x^{2}}\tanh{x}\\
    \approx 4\pi \nu r_{s} k_{F}d \ln\frac{1}{k_{F}d}.
\end{multline}
Comparing to the interaction at small momenta $\nu U(q\to 0) = r_{s}k_{F}d$ we extract the effective interaction strength and cutoff
\begin{equation}
    \nu U \sim  r_{s} k_{F} d,\quad \Lambda \sim \frac{1}{d}.
\end{equation}

\end{document}